\documentclass{WileyMSP-template}
\usepackage{hyperref}% add hypertext capabilities
\usepackage{siunitx}
\usepackage{amsmath}
\usepackage{multirow}
\hypersetup{
    colorlinks=true,
    linkcolor=blue,
    filecolor=magenta,      
    urlcolor=cyan,
    citecolor = green,
    pdftitle={Overleaf Example},
    pdfpagemode=FullScreen,
    }
\newcommand{\pd}[2]{\frac{\partial #1}{\partial #2}}
\begin{document}

\pagestyle{fancy}
\rhead{\includegraphics[width=2.5cm]{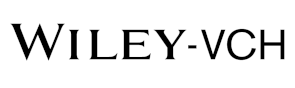}}

\title{Multiphase  polarization in ion-intercalation nanofilms: general theory including various surface effects and memory applications}
%Multiphase concentration polarization in ion-intercalation materials: general theory and prospects for memory applications

\maketitle

% Author: Please give full first and last names for authors and include * after the name of all corresponding authors

\author{Huanhuan Tian}
\author{Ju Li}
\author{Martin Z. Bazant*}

% Dedication

%\dedication{Optional dedication here. If no dedication is required, please leave blank}

% Affiliations: Please provide adacemic titles (Prof. or Dr.) for all authors where applicable, and include an institutional email address for all corresponding authors
\begin{affiliations}
Huanhuan Tian \\
Department of Chemical Engineering, Massachusetts Institute of Technology, Cambridge, MA 02139 

Prof. Ju Li\\
Department of Nuclear Science and Engineering and Department of Materials Science and Engineering, Massachusetts Institute of Technology, Cambridge, MA 02139 

Prof. Martin Z. Bazant\\
Department of Chemical Engineering and Department of Mathematics, Massachusetts Institute of Technology, Cambridge, MA 02139\\
Email Address: bazant@mit.edu

\end{affiliations}

% Keywords: Please provide a minimum of three and a maximum of seven keywords, separated by commas


% Abstract should be written in the present tense and impersonal style (i.e., avoid we), and be at most 200 words long
\begin{abstract}
% this version has 191 words
%The multiphase, coupled ion-electron transport can be very important in ion-intercalation materials. For example, in multiphase ion-intercalation nanofilms sandwiched by ion-blocking electrodes, ion concentration polarization (current-induced concentration gradient near a charge-selective interface)  above certain threshold currents can change phase distribution, as predicted by a preliminary 1D model in our recent publication. The interfacial phase change (IPC) on electrodes can further lead to interfacial resistive switching (RS) for asymmetric electrodes with ion-modulated electron transfer, which can be utilized to make nonvolatile RS memories for in-memory computing.  In this work, we derive a comprehensive 2D phase-field model for coupled ion-electron transport in ion-intercalation materials, with surface effects including electron transfer kinetics, non-neutral wetting, energy relaxation, and surface charge. Then we use the model to study IPC. We present various patterns for time evolution of phase boundaries at different conditions, and show that the surface heterogeneity, surface charge, and non-neutral wetting can reduce the switching current significantly. In addition, we compare the physics and performance of IPC with other non-volatile RS mechanisms, and show that IPC-based memories require multiphase ion-intercalation materials with high ionic diffusivity, low electronic conductivity, and significant metal-insulator transition with concentration.

% this version has 188 words (max: 200)
Ion concentration polarization (CP, current-induced concentration gradient adjacent to a charge-selective interface) has been well studied for single-phase mixed conductors (e.g., liquid electrolyte), but multiphase CP has been rarely addressed in literature. In our recent publication, we proposed that CP  above certain threshold currents can flip the phase distribution in multiphase ion-intercalation nanofilms sandwiched by ion-blocking electrodes. We call this phenomenon as multiphase polarization (MP). We then proposed that MP can further lead to nonvolatile interfacial resistive switching (RS) for asymmetric electrodes with ion-modulated electron transfer, which theory can reproduce the experimental results of LTO memristors.  In this work, we derive a comprehensive 2D phase-field model for coupled ion-electron transport in ion-intercalation materials, with surface effects including electron transfer kinetics, non-neutral wetting, energy relaxation, and surface charge. Then we use the model to study MP. We present time evolution of phase boundaries, and analyze the switching time, current, energy, and cyclic voltammetry, for various boundary conditions. We find that the switching performance can be improved significantly by manipulating surface conditions and mean concentration. Finally, we discuss the prospects of MP-based memories and possible extensions of the current model. 

\end{abstract}

% Text: Please use section headings and subheadings as specified below. For communications, all section headings apart from Experimental Section should be removed
% Please make the first reference to a display item bold: \textbf{Figure 1}
% Do not abbreviate Figure, Equation, etc.; display items are always singular, i.e., Figure 1 and 2.
% Equations are always singular, i.e., Equation 1 and 2, and should be inserted using the {equation} environment, not as graphics
% Please do not use footnotes in the text, additional information can be added to the Reference list.

\section{Introduction}
Ion-intercalation materials, which allow reversible insertion of host ions (along with electrons) and ion-modulation of certain physical properties (e.g., thermodynamic, electronic, optical, and magnetic properties), play important roles in various applications such as energy storage \cite{Nitta2015Li-ionFuture}, ion separation \cite{Singh2018TheoryMaterials, Singh2019TimelineDeionization}, electrochromic display \cite{Giannuzzi2015OnDevices}, and very recently, information storage and computing \cite{Gonzalez-Rosillo2020Lithium-BatterySeparation, Onen2022NanosecondLearning}. Many commonly used ion-intercalation materials \cite{Nitta2015Li-ionFuture} (e.g.,  Li$_{4+3x}$Ti$_5$O$_{12}$ (LTO) \cite{Zhao2015APerspectives},  Li$_x$FePO$_4$ (LFP) \cite{Cogswell2012CoherencyNanoparticles}) exhibit multiphase behaviours and mixed ion-electron conductivity. Therefore, it is important to understand the coupled ion-electron transport within and at the interfaces of multiphase ion-intercalation materials, especially at high electric currents during rapid operations \cite{Onen2022NanosecondLearning, Liu2019ChallengesMaterials}, but related studies are still limited.  
%tungsten oxide \cite{Dini1996ANa+, Zhong1992LithiumLixWO3},

The multiphase ion-intercalation materials can spontaneously split into ion-rich and ion-poor phases  if the mean ion concentration falls in the spinodal region, where co-existing phases are thermodynamically more stable than single phase \cite{Bazant2017ThermodynamicElectro-autocatalysis}. Numerous phase-field models have been developed to study the multiphase ion intercalation processes and have brought lots of insights to applications such as Li-ion batteries \cite{Han2004ElectrochemicalModels, Bazant2013TheoryThermodynamics, Dreyer2010TheBatteries,  Bai2011SuppressionDischarge, Cogswell2012CoherencyNanoparticles}. However, most phase-field models do not consider coupled ion-electron transport \cite{Han2004ElectrochemicalModels, Bazant2013TheoryThermodynamics, Dreyer2010TheBatteries, Bai2011SuppressionDischarge, Cogswell2012CoherencyNanoparticles}. They first solve the transport of neutral ion-electron pairs during intercalation, which is a pure diffusion problem with a flux boundary (Cahn-Hilliard reaction model \cite{Bazant2013TheoryThermodynamics}) or a diffusion-reaction problem without boundary flux (Allen-Cahn model \cite{Bazant2013TheoryThermodynamics}). And then they calculate the ion-modulated electronic conductance if needed \cite{Fraggedakis2020, Nadkarni2019ModelingStorage}. This simplification works well if the electronic conductivity is much larger than ionic conductivity and the applied current is not too large,  regardless of the multiphase behaviors  \cite{Wepper1978ElectrochemicalSolids, Maier2004PhysicalSolids}. Otherwise, the coupling of ion and electron transport may not be neglected.

Much research has been done for the coupling of ion and electron transport within and at interfaces of single-phase ion-electron conductors \cite{Wepper1978ElectrochemicalSolids, Maier2004PhysicalSolids,Guyer2004PhaseEquilibrium, Guyer2004PhaseKinetics}. For example, it is known that high electronic currents can lead to ion concentration polarization (CP, ion enrichment on one electrode and depletion on the other) in a mixed ion-electron conductor between ion-blocking electrodes. This CP can influence the measurement of electronic conductivity \cite{Fehr2010DCSpinel}, and lead to some volatile memristive behaviors of semiconductors \cite{Strukov2009CoupledBehavior}. CP has also been well studied in liquid electrolytes and has been utilized for water treatment \cite{Mani2011DeionizationMicrostructures, Dydek2013NonlinearModel, Tian2021ContinuousElectrodialysis}. 
%Another example of coupled ion-electron transport is the significant impact of the space charge composed of ions and electrons on heterogeneous reaction at high currents \cite{Guyer2004PhaseEquilibrium, Guyer2004PhaseKinetics}. %formation of space charge composed of ions and electrons at interfaces with potential drop \cite{Mebane2015ASolutions, Zurhelle2017ASolution}. For instance, the impact of the space charge on heterogeneous reaction at high currents can be significant \cite{Guyer2004PhaseEquilibrium, Guyer2004PhaseKinetics}. 
%In addition, phase-field models have also been developed for coupled ion-electron transport near electrochemical interfaces between electrolyte and electrodes \cite{Guyer2004PhaseEquilibrium, Guyer2004PhaseKinetics}. 
However, coupled ion-electron transport such as CP for multiphase ion-intercalation materials remain poorly addressed in the literature.

\begin{figure*}[hbt!]
    \centering
    \includegraphics[width = \textwidth]{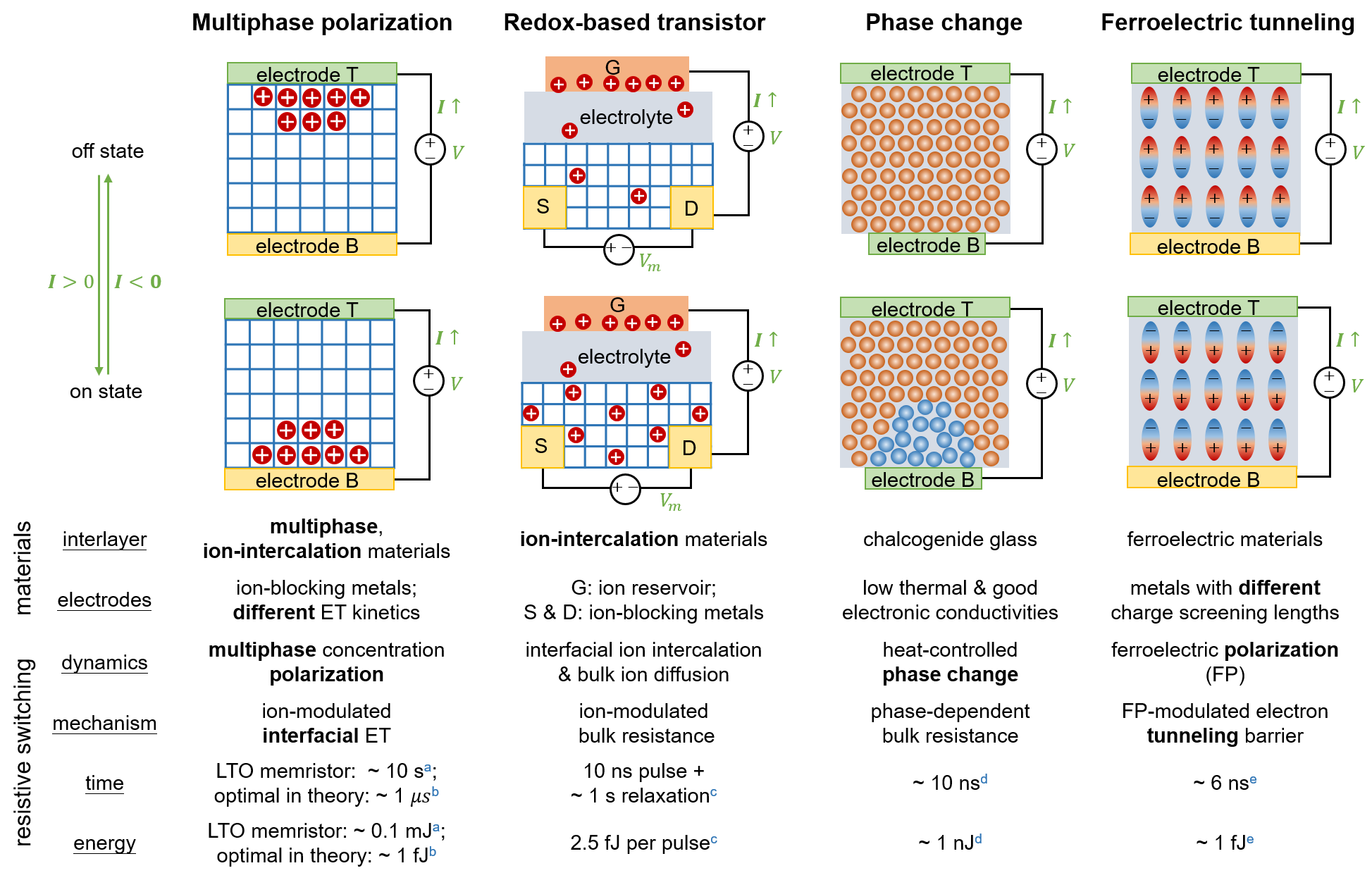}
    \caption{Comparison between four nonvolatile resistive switching mechanisms: multiphase polarization (MP, this work), redox-based transistor (RBT), phase change (PC), and ferroelectric funneling (FT). ET represents for electron transfer. Only single crystals are drawn for simpicity, but polycrystals are common in real devices. References: a, estimated from Figure 2(c) in Ref. \cite{Gonzalez-Rosillo2020Lithium-BatterySeparation} (see the calculation in \autoref{sec:switching_cur_time_c0.1}); b, prediction  in this work (\autoref{sec:discussion}); c, Ref.\cite{Onen2022NanosecondLearning}; d, Ref.\cite{Ding2019Phase-changeOperation}; e, Ref.\cite{Huang2018DiffusionBeyond}.}
    \label{fig:mechanism}
\end{figure*}

In our recent work \cite{Tian2022InterfacialNanofilms}, we described the CP in multiphase ion-intercalation materials between ion-blocking electrodes, based on a 1D preliminary phase-field model for coupled cation-electron transport. At high electric currents, multiphase CP can lead to phase redistribution, a phenomenon we call multiphase polarization (MP).   As shown in the first panel of \autoref{fig:mechanism}, a large enough downward electric current in the nanofilm can drive the ion-rich phase to the bottom electrode or drive the ion-poor phase to the top electrode, and an upward current can drive the opposite process. Unlike CP which disappears at zero currents, MP is nonvolatile since the altered phase distribution remains even after removal of the currents. We then assumed that interfacial electron transfer resistance dominates the total resistance and strongly depends on local ion concentration, therefore MP-induced interfacial phase change (IPC) can lead to non-volatile interfacial resistive switching (RS) if the two electrodes are asymmetric. We then gave a thorough analysis of the switching time and current, resistance ratio, and cyclic voltammetry behaviors for the MP mechanism using a 1D preliminary model with natural boundary condition (zero normal gradient of concentration), and  qualitatively reproduced the experimental RS behaviors of LTO memristors (made from LTO nanofilm sandwiched by Pt electrodes) \cite{Gonzalez-Rosillo2020Lithium-BatterySeparation}. %In this work, we emphasize the concept of IPC instead of MP, to address the ``interfacial" aspect. We also call the nonvolatile RS mechanism described in Ref.\cite{Tian2022InterfacialNanofilms} as IPC. However, 
Note that  MP itself is a more general concept, while MP-based RS has more requirements for the ion-intercalation materials and electrodes (more discussion can be found in \autoref{sec:discussion}). 

Non-volatile RS devices can be used for  in-memory computing, which revolutionizes traditional computing architectures by solving the von Neumann bottleneck problem \cite{Ielmini2018In-memoryDevices}. The two-terminal non-volatile RS devices are usually called memristors. We can then compare MP with other well-known and relevant non-volatile RS mechanisms \cite{Ielmini2018In-memoryDevices, Wang2020ResistiveProcessing}, including redox reactions \cite{Sawa2008ResistiveOxides, Waser2009Redox-basedChallenges, Valov2011ElectrochemicalProspects, Ielmini2016ResistiveScaling,   DelValle2018ChallengesComputing},  phase change \cite{ Ding2019Phase-changeOperation, Wong2010PhaseMemory, DelValle2018ChallengesComputing, Simpson2011InterfacialMemory}, and ferroelectric tunneling \cite{Garcia2014FerroelectricProcessing, Huang2018AJunctions, Mikolajick2020TheMemories, Setter2006FerroelectricApplications}, as shown in \autoref{fig:mechanism}.  Redox-based mechanisms usually change the bulk resistance  of a nanofilm by injecting charge carriers through an active electrode \cite{Sawa2008ResistiveOxides, Waser2009Redox-basedChallenges, Valov2011ElectrochemicalProspects, Ielmini2016ResistiveScaling, Wang2020ResistiveProcessing, Ielmini2018In-memoryDevices, DelValle2018ChallengesComputing}. Many redox-based memristors involves the formation of conductive filaments \cite{Ielmini2018In-memoryDevices, Wang2020ResistiveProcessing}. Recently, great progress has been made in filament-free redox-based transistors (RBT) using ion-intercalation materials with ion-modulated conductivity \cite{Fuller2017Li-IonComputing, Yao2020ProtonicNetworks, Onen2021CMOS-CompatibleLearning,  Onen2022NanosecondLearning}. Compared with RBT, MP has conserved total amount of ions during switching and mainly changes surface resistance instead of bulk resistance. The so-called phase change (PC) mechanism usually uses Joule heating to switch between amorphous and crystalline phases which have different bulk resistance \cite{  Ding2019Phase-changeOperation, Wong2010PhaseMemory,Ielmini2018In-memoryDevices, Wang2020ResistiveProcessing}, while MP uses electric field to move  ion-modulated phases which can change interfacial resistance. In the ferroelectric tunneling (FT) mechanism, the ferroelectric polarization switchable by a threshold voltage can change the tunneling current if the two electrodes have different charge screening length  \cite{Wang2020ResistiveProcessing, Ielmini2018In-memoryDevices, Mikolajick2020TheMemories, Setter2006FerroelectricApplications}. Interestingly, MP and FT are in some sense conceptually similar, since both of them utilize some non-volatile polarization of the bulk materials and the asymmetry of the electrodes to change the total resistance.  In \autoref{sec:discussion},   we will make more comparison about the performance of these mechanisms.

In this work, we aim to develop a comprehensive and general phase field model for coupled, multiphase ion-electron transport with electron transfer kinetics, non-neutral wetting, energy relaxation and electric double layer at the electrode-nanofilm interfaces. To the best of our knowledge, this model is the first non-equilibrium model to consider surface energy relaxation and surface charge of multiphase ion-intercalation materials. Then we use the new model to analyze 2D MP with various surface effects included, and present discussion in the general outlook. 
%This work can better quantitatively explain the LTO memristor experiments \cite{Gonzalez-Rosillo2020Lithium-BatterySeparation} and provide more insights for device design. 

The paper is organized as below. We first present the derivation of the phase-field model in \autoref{sec:model}, and then do some thermodynamic and kinetic analysis of the model in \autoref{sec:analysis}. In \autoref{sec:eqs}, we apply the general model to regular solutions to model ion-intercalation material, and list all the dimensionless equations to be solved. We then simulate 2D MP at various conditions, and compare the results with the old 1D model in \autoref{sec:results}. Finally, we discuss the prospects of MP-based memristors and the model itself, and give a conclusion in \autoref{sec:discussion}.

\section{General phase-field model}{\label{sec:model}}
\subsection{Introduction}
\begin{figure}
    \centering
    \includegraphics[width = 0.45 \columnwidth]{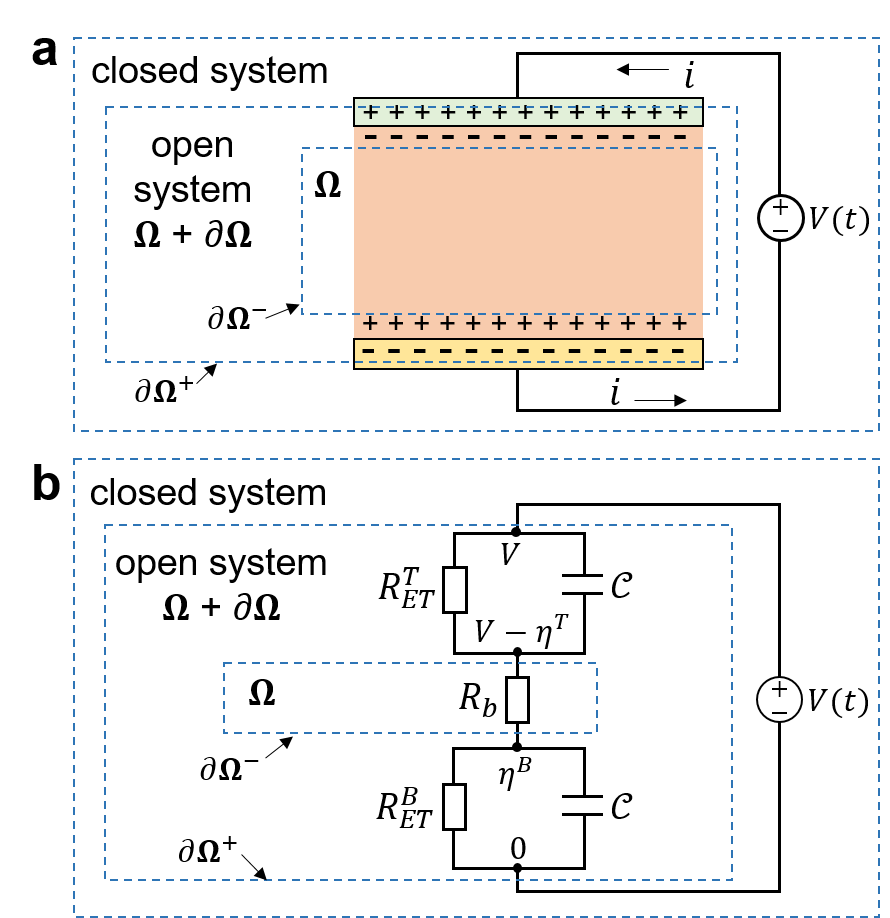}
    \caption{ (a) The model system that we consider in this work and (b) its equivalent circuit model when the interlayer is homogeneous. }
    \label{fig:system}
\end{figure}

In this section, we derive a general phase-field model for multiphase coupled ion-electron transport in a nanofilm with thickness $L$ sandwiched between two parallel ion-blocking electrodes, as shown in \autoref{fig:system}(a). We expect that a voltage applied to the electrodes can result in electron transfer and charge accumulation at the electrode-nanofilm interfaces, as well as phase re-distribution in the nanofilm.  Following our previous work \cite{Tian2022InterfacialNanofilms}, we consider three species in the mixed ion-electron conductor: monovalent cations ``p", electrons ``n", and immobile positively charged defects ``d". We assume that the mobile cations and electrons are fully dissociated and ignore the generation and combination of charge carriers.  We also assume constant pressure and temperature, and ignore mechanical effects and volume change (a good assumption for “zero strain” materials like Li$_{4+3x}$Ti$_5$O$_{12}$ \cite{Zhao2015APerspectives} and Li$_x$WO$_3$\cite{Zhong1992LithiumLixWO3}). In addition, we assume that the total resistance is dominated by surface resistance thus the potential drop in the bulk (usually within several thermal voltages in each phase) is very small compared with the total potential drop (up to hundreds of thermal voltages). Furthermore, we consider large ion concentration (thousands of moles per cubic meter, which is typical for ion-intercalation materials) so that the space charge should be very thin compared with the nanofilm (10-100 nm).   

%We also assume the total resistance is dominated by surface resistance (electron transfer resistance), following our previous work \cite{Tian2022InterfacialNanofilms}.

Then we can divide the open system shown in \autoref{fig:system}(a) into two parts: the bulk domain $\mathbf\Omega$, and the surface $\partial \mathbf{\Omega}$. We assume no electrostatic energy everywhere in $\mathbf\Omega$ (which can lead to local electroneutrality), and put the electric double layer (EDL, formed by charge from the nanofilm and the electrode) with net neutrality into $\partial \mathbf{\Omega}$. We ignore the diffuse part of EDL since we consider large ion concentration. In addition, we ignore the EDL at phase boundaries (interfaces between different phases), whose contribution to phase boundary energy (and potential drop) should not change much with currents  due to the assumed small bulk resistance. In another word, we can just control the effective phase boundary energy by adjusting some parameters (e.g., $\kappa_p$ in \autoref{sec:interface}) without modeling the EDL explicitly at phase boundaries.  The above treatment greatly simplifies the problem  and can provide lots of insights.  Then the total Gibbs free energy of the open system can be expressed as: 
\begin{equation}
    G (c_p, c_n, \phi) = G_b + G_s = \int_{\mathbf{\Omega}} g dV  + \int_{ \partial \mathbf{\Omega}} \gamma_s  dS,
    \label{eq:G}
\end{equation}
where $c_p$ and $c_n$ are the molar concentration of the cations and the electrons, $\phi$ is the electric potential, $g$ is the local free energy density in the bulk $\mathbf{\Omega}$, and $\gamma_s$ is the surface energy density on $\partial \mathbf{\Omega}$. 

%In this section, we want to derive the governing differential equations and boundary conditions for the non-equilibrium (kinetic) system. 

Next, based on the first and second laws of thermodynamics, the total Gibbs free energy $\mathcal{G}$ of the closed system  in \autoref{fig:system} at constant pressure and temperature should not increase with time \cite{Bejan2016AdvancedThermodynamics}. If we assume no energy dissipation in the voltage source, we have 
\begin{equation}
    \pd{\mathcal{G}}{t} = \pd{G}{t} + \int_{\partial \mathbf{\Omega}^+} \mu_n \mathbf{J}_n \cdot \mathbf{n} dS \leq 0,
\end{equation}
where the integral is the energy flux out of the open system (or the energy flux into the voltage source), $\mu_n$ is the electrochemical potential of electrons,  $\mathbf{J}_n$ is the flux of electrons, and $\mathbf{n}$ is the normal vector on $\partial \mathbf{\Omega}$ to point outward of $\mathbf \Omega$. Noting that the energy flux into and out $\partial \mathbf{\Omega}$ may not be the same, we denote $\partial \mathbf{\Omega}^+$ and $\partial \mathbf{\Omega}^-$ as the outer (electrode) and inner (nanofilm) side of $\partial \mathbf{\Omega}$, respectively. For the convenience of mathematical derivation, we can further split $\pd{\mathcal{G}}{t}$ into two parts $\pd{\mathcal{G}}{t} = \pd{\mathcal{G}_1}{t} + \pd{\mathcal{G}_2}{t}$:
\begin{equation}
    \pd{\mathcal{G}_1}{t} = \pd{G_b}{t} + \int_{\partial \mathbf{\Omega}^-}  \mu_n \mathbf{J}_n \cdot \mathbf{n} dS,
\end{equation}
\begin{equation}
    \pd{\mathcal{G}_2}{t} = \pd{G_s}{t} + \int_{\partial \mathbf{\Omega}^+} \mu_n \mathbf{J}_n \cdot \mathbf{n} dS - \int_{ \partial \mathbf{\Omega}^-} \mu_n \mathbf{J}_n \cdot \mathbf{n} dS.
    \label{eq:dG2dt_def}
\end{equation}

In the following of this section, we derive $\pd{\mathcal{G}_1}{t}$ and the corresponding bulk equations in \autoref{sec:model_bulk}, derive $\pd{\mathcal{G}_2}{t}$ and the corresponding  boundary conditions in \autoref{sec:model_surf}, and give a summary and discussion in \autoref{sec:model_sum}.

\subsection{Bulk energy evolution}{\label{sec:model_bulk}}

In this part, we derive $\pd{\mathcal G_1}{t}$ and the corresponding bulk governing equations.  In $\mathbf \Omega$, we assume the local free energy density $g$ is just the summation of the free energy density for each species 
\begin{equation}
    g = g_p + g_n + g_d.
\end{equation}
We then assume the fixed charge ``d'' only has electrostatic energy,
\begin{equation}
    g_d = z_d c_d F \phi,
\end{equation}
and the mobile species    ``p'', ``n'' have chemical energy, electrostatic energy, and Cahn-Hilliard gradient energy \cite{Cahn1958FreeEnergy, Bazant2013TheoryThermodynamics, Guyer2004PhaseEquilibrium, Mebane2015ASolutions}:
\begin{equation}
    g_k(c_k, \phi) = g_{k, chem} (c_k) + z_k c_k F \phi + \frac{1}{2}\kappa_k \left|\nabla c_k \right|^2,
\end{equation}
where $z_k$ is the valence of species $k$, $F$ is the Faraday constant, $\phi$ is the electric potential, and $\kappa_k$ is the diagonal gradient energy coefficient (assumed to be a constant). In this work, we further assume $\kappa_n = 0$, which can help us simplify the expression for surface energy in \autoref{sec:model_surf}. Note that here we have neglected the electrostatic energy term $-\frac{1}{2}\epsilon|\nabla \phi|^2$ in $g$ ($\epsilon$ is the permittivity), which can lead to space charge \cite{Guyer2004PhaseEquilibrium, Mebane2015ASolutions} and dielectric breakdown (if $\epsilon$ is a nonlinear function of ion concentration) \cite{Fraggedakis2020, Chen2008Phase-fieldReview, ChaitanyaPitike2014Phase-fieldSolids} at high electric fields. This treatment is reasonable in this work, since we have assumed small bulk electric fields (within a few thermal voltages over 100 nm) and thin space charge (due to the assumed large concentration), and have put space charge into $\partial \mathbf \Omega$. In addition, since we ignore the generation-recombination of electrons and ions, the conservation of species leads to
\begin{equation}
    \pd{c_k}{t} + \nabla \cdot \mathbf{J}_k = 0,
\end{equation}
where $\mathbf{J}_k$ is the flux of species $k$.

Now after some algebra (see Appendix A),  we get
\begin{equation}
\begin{aligned}
     \pd{\mathcal{G}_1}{t} &=  \sum_{k=p,n}\int_{\mathbf{\Omega}} \nabla \mu_k \cdot \mathbf{J}_k   dV   \\
     +&  \int_{\mathbf{\Omega}} \left(\sum_{k=p,n,d}z_k c_k F  \right)\pd{\phi}{t} dV  \\
     +& \int_{\partial \mathbf{\Omega}^-}  \kappa_p \mathbf{n} \cdot \nabla c_p   \pd{c_p}{t} dS,
\end{aligned}
\label{eq:dG1dt}
\end{equation}
where  $\mu_k$ is the electrochemical potential of species $k$ in the nanofilm \cite{Bazant2013TheoryThermodynamics}
\begin{equation}
\begin{aligned}
        \mu_k &= \pd{g_k}{c_k} - \nabla \cdot \pd{g_k}{\nabla c_k} \\
        &= \mu_{k,chem} + z_k F \phi - \kappa_k \nabla^2 c_k ,
    \label{eq:mu_k}
\end{aligned}
\end{equation}
where $\mu_{k,chem} = \pd{g_{k, chem}}{c_k}$ is the chemical potential.
The surface integral in \autoref{eq:dG1dt} can be combined with $\pd{\mathcal G_2}{t}$ to get the boundary conditions in \autoref{sec:model_surf}.  Now we can get the bulk governing equations that ensures the volume integrals in \autoref{eq:dG1dt}  minimize spontaneously 
\begin{equation}
    \mathbf{J}_k = -\frac{ D_k c_k}{RT} \nabla \mu_k,
    \label{eq:flux}
\end{equation}
\begin{equation}
    c = c_p =  -( z_n c_n + z_d c_d)/z_p,
    \label{eq:e_neutral}
\end{equation}
where  $D_k$ is the diffusivity, $R$ is the gas constant. The corresponding conductivity is $\sigma_k = D_k F^2 z_k^2 c_k / R T$. \autoref{eq:flux} assumes that the flux is linear with the gradient of $\mu_k$ which should work well for near-equilibrium dynamics. \autoref{eq:e_neutral} is the electroneutrality equation, which reduces the two variables $c_p$ and $c_n$ into one variable $c$.  Taking $z_p = 1$, $z_n = -1$, $z_d = 1$, we can reduce \autoref{eq:e_neutral} to
\begin{equation}
    c  = c_p = c_n - c_d.
    \label{eq:e_neutral2}
\end{equation}

\subsection{Surface energy evolution}{\label{sec:model_surf}}

%This also simplifies $\sum_{k=p, n} \kappa_k \mathbf{n} \cdot \nabla c_k \pd{c_k}{t} = \kappa \mathbf{n} \cdot \nabla c\pd{c}{t}$  where $\kappa = \kappa_p  - \kappa_n z_p/z_n$. 

Next we derive $\pd{\mathcal{G}_2}{t}$ and get the boundary conditions.  We first need to zoom in to see the internal structure of the surface. Inside $\partial \mathbf{\Omega}$, we assume there is a net-neutral EDL composed of parallel, infinitely-thin, and opposite charges that are separated by a small distance $d_s$ and formed by excess/lack of electrons on the surfaces of the nanofilm and electrode, respectively. This indicates that the diffuse part and the ionic component of the nanofilm surface charge are both neglected. 
%We also assume net electroneutrality in $\partial \mathbf{\Omega}$, so that $\mathbf{J}_n$ on $\partial \mathbf{\Omega}^-$ and $\partial \mathbf{\Omega}^+$ are the same. 
In the following, we use $\Delta (\cdot)$ to represent the jump of a surface variable across $\partial \mathbf{\Omega}$ (the variable on the electrode surface minus that on the nanofilm surface). Then we can model the EDL as a linear capacitor with $\Delta \phi$-independent capacitance density $C = \epsilon_s/d_s$ ($\epsilon_s$ is the permittivity between the two charged surfaces), and get
\begin{equation}
    \Gamma = C \Delta \phi,
\end{equation}
where $\Gamma$ is the surface charge density  on the electrode side of the EDL.  This linear capacitor without ionic charge greatly simplifies the problem. More complex surface charge models can be considered in future works (see discussion in \autoref{sec:discussion}). We can further split $\Delta \phi$ into two parts by defining the overpotential $\eta$
\begin{equation}
    \eta =  \frac{1}{z_n F}\Delta \mu_n = \Delta \phi - (\Delta \phi)_{0},
    \label{eq:eta}
\end{equation}
where $(\Delta \phi)_{0}$ is only determined by chemical part of $\Delta \mu_n$ and thus only depends on $c$ but not $\eta$.

Then we want to derive an expression for surface energy density $\gamma_s$, which should only be a function of overpotential $\eta$ and nearby $c$ on $\partial \mathbf \Omega^-$ (remember that the bulk electroneutrality condition, \autoref{eq:e_neutral2}, has reduced $c_p$ and $c_n$ to a single parameter $c$). 
%First, we denote the surface energy density at zero overpotential as $\gamma_s^0$, which should only depend on $c$.  We will discuss  $\gamma_s^0(c)$ in detail in \autoref{sec:interface}.  We then want to derive the excess surface energy density obtained from overpotential ($\gamma_s - \gamma_s^0$, which equals zero if $\eta=0$). 
Here we present two approaches. 

In the first approach, we notice three contributions of $\gamma_s$: the free energy of the surface charges, the electrostatic energy between the surface charges, and the structural energy (non-electric part of the surface energy that may depend on $c$). Since we have assumed $\kappa_n=0$, $\Delta \mu_n$ and $\gamma_s$ should not depend on $\nabla c$. Therefore, we have
\begin{equation}
    \gamma_s(c, \eta) = \Delta \mu_n \frac{\Gamma}{z_n F} - \frac{1}{2}\epsilon_s \left(\frac{\Delta \phi}{d_s} \right)^2 d_s + \gamma_s^*(c),
    \label{eq:gamma_s_ap1_step1}
\end{equation}
where we have assumed constant electric field between surface charges, and $\Gamma$-independent chemical potential of electrons at the charged surfaces.  Then we can substitute \autoref{eq:eta} into \autoref{eq:gamma_s_ap1_step1}, and rearrange \autoref{eq:gamma_s_ap1_step1} to get
\begin{equation}
    \gamma_s(c, \eta) = \frac{1}{2}C\eta^2 + \gamma_s^0(c),
    \label{eq:gamma_s_ap1}
\end{equation}
where $\gamma_s^0$ is the surface energy density at zero overpotential. Note that $\gamma_s^*$ does not equal $\gamma_s^0$ unless $(\Delta\phi)_0 = 0$.

In the second approach, we integrate the work done to charge the surfaces as the surface potential drop increases from $(\Delta \phi)_0$ to $\Delta \phi$ \cite{Verwey1948TheoryColloids} 
\begin{equation}
    \gamma_s(c, \eta)  - \gamma_s^0 (c) =  \int_{C (\Delta \phi)_0}^{C \Delta \phi} \Delta \mu_n \frac{d\Gamma}{z_n F}  = \frac{1}{2}C\eta^2.
    \label{eq:gamma_s_ap2}
\end{equation}

As we can see, we get the same results from the two approaches. The excess surface energy $\gamma_s - \gamma_s^0$ is the same as the energy stored in a capacitor at potential drop $\eta$. More details for discussion of the signs of the surface energy can be found in Appendix B. 

Now we can put $\gamma_s$ into \autoref{eq:dG2dt_def} and get  
\begin{subequations}
\begin{equation}
\begin{aligned}
    \pd{\mathcal{G}_2}{t}   = \int_{\partial \mathbf{\Omega}} \pd{\gamma_s^\mathrm{eff}}{c} \pd{c}{t}dS  -  \int_{\partial \mathbf{\Omega}}  F\eta  \mathbf{J}_n^{ET} \cdot \mathbf{n} dS ,
    \label{eq:dG2dt}
\end{aligned}
\end{equation}
\begin{equation}
     \pd{\gamma_s^\mathrm{eff}}{c} = \pd{\gamma_s^0}{c} - C \eta \pd{(\Delta \phi)_0}{c} + \left(\frac{1}{2}\eta^2 - \eta \Delta \phi \right)\pd{C}{c}.
     \label{eq:dgamma_eff_dc}
\end{equation}
\end{subequations}
Here we have used $\left(\mathbf{J}_n \cdot \mathbf{n} \right)_{\partial \mathbf{\Omega}^\pm} = \mathbf{J}_n^{ET} \cdot \mathbf{n} + \frac{1}{F}\pd{C \Delta \phi}{t}$ from mass balance for surface charge, where $\mathbf{J}_n^{ET} \cdot \mathbf{n}$ is the electron transfer rate between the two charged surfaces as a function of $c$ and $\eta$. $\gamma_s^{\mathrm{eff}}$ has included energy change of the voltage source compared with $\gamma_s$.   In addition, if we assume the chemical potential of electrons in electrodes is a constant, then we can plug $\pd{\Delta \phi_0}{c} = \frac{1}{z_n F}\pd{\mu_{n, chem}}{c}$ into \autoref{eq:dgamma_eff_dc} to make it more explicit. In the following, for simplicity, we also assume $C$ does not depend on $c$, though this dependence can be easily added in future works.

Finally, we add $\pd{\mathcal G_1}{t}$ and $\pd{\mathcal G_2}{t}$ together, and let the surface integrals also not increase with time. This can be satisfied by applying  the following conditions 
\begin{equation}
\begin{aligned}
    \pd{\gamma_s^\mathrm{eff}}{c}  + \kappa_p \nabla c \cdot \mathbf{n} = -\tau_\gamma \pd{c}{t},
    \label{eq:gamma_BC}
\end{aligned}
\end{equation}
\begin{equation}
    \mathbf{n} \cdot \mathbf{J}_n^{ET} \eta >0.
    \label{eq:Jn_eta_neq}
\end{equation}
In \autoref{eq:gamma_BC}, $\tau_\gamma$ (a positive number) adds relaxation to surface energy and can lead to dynamic contact angle \cite{Qian2003MolecularFlows, Yue2011WallLines, Qian2006AHydrodynamics, Bonn2009WettingSpreading}, and  $\pd{\gamma_s^{\mathrm{eff}}}{c}$  instead of $\pd{\gamma_s}{c}$ determines steady state surface energy balance (as well as contact angle). We discuss on \autoref{eq:gamma_BC} in detail in \autoref{sec:contact_angle}.    \autoref{eq:Jn_eta_neq} simply requires that the electrons should transfer against the overpotential, which is consistent with the existing electron transfer models like Bulter-Volmer equation \cite{Bazant2013TheoryThermodynamics, Tian2022InterfacialNanofilms}, Marcus theory \cite{Marcus1964ChemicalTheory}, and Schottky diode \cite{Sze2007PhysicsDevices}.  

\subsection{Summary and discussion }{\label{sec:model_sum}}

Based on \autoref{sec:model_bulk} and \autoref{sec:model_surf}, now we can write the time evolution of the total free energy of the closed system as
\begin{equation}
\begin{aligned}
     \pd{\mathcal{G}}{t} &=  \int_{\mathbf{\Omega}} \sum_{k=p,n} \nabla \mu_k \cdot \mathbf{J}_k   dV   \\
     +& \int_{\partial \mathbf{\Omega}} \left(\pd{\gamma_s^\mathrm{eff}}{c}  + \kappa_p \nabla c \cdot \mathbf{n} \right) \pd{c}{t} \\
     -&  \int_{\partial \mathbf{\Omega}} \eta F \mathbf{J}_n^{ET} \cdot \mathbf{n} dS,
     \label{eq:dGdt}
\end{aligned}
\end{equation}
where the three integrals represent the energy dissipation by bulk transport, surface friction, and interfacial electron transfer, respectively. The energy transfer between the voltage source and surface charges does not change the total free energy.

If we assume homogeneous single phase in the nanofilm,  we can analogize the whole system as an electric circuit shown in \autoref{fig:system}(b). In this case, the second integral in \autoref{eq:dGdt} disappears. The integrand in the first volume integral becomes
\begin{equation}
   \sum_{k=p,n} \nabla \mu_k \cdot \mathbf{J}_k \rightarrow - \sigma |\nabla \phi|^2
\end{equation}
where $\sigma = \sum_{k=p,n} \sigma_k = \sum_{k=p,n} z_k^2 F^2 c_k D_k/RT $ is the bulk conductivity. Therefore, the first integral in \autoref{eq:dGdt} represents the energy dissipation on a bulk resistance $R_b$. Similarly, the last integral in \autoref{eq:dGdt} represents the energy dissipation on  two surface resistances $R_{ET}^T$ and $R_{ET}^B$ (T: top electrode, B: bottom electrode).  In addition, the energy can be transferred between the voltage source and the two surface charge capacitors at electrode-nanofilm interfaces. Each capacitor has capacitance  $\mathcal{C} = CA$ ($A$ is surface area) and potential drop $\eta$.  Finally, the RC circuit is connected to a voltage source $V(t)$ to close the system. We can see clearly that in the circuit, the energy input from the voltage source 
has two destinations: dissipated in the resistances $R_b$, $R_{ET}^T$ and $R_{ET}^B$, or stored in the capacitors.

In this work, we care more about the multiphase behaviours in the nanofilm. In this case, the system is more complex than the circuit model shown in \autoref{fig:system}(b). We assume ion concentration can significantly influence the electron transfer rate at electrode-nanofilm interfaces.  Therefore, MP-induced IPC at high electric currents can change the resistance $R_{ET}^T$ and $R_{ET}^B$, leading to RS behaviors.

\section{Thermodynamic and kinetic analysis}{\label{sec:analysis}}
In this section, we derive some analytical expressions based on the phase field model, including the phase equilibrium properties, the critical current for MP (switching current), and the surface energy and contact angle. Finally, we compare the surface charge effects in our system with the well-known electrowetting phenomenon. 

\subsection{Phase equilibrium}
In this part, we analyze the equilibrium state (no flux everywhere) and derive the binodal points, spinodal points, and equilibrium contact angle. To begin with, we define $g_h$ as the homogeneous part of the total free energy density $g$,  and define $\mu_h$ as the homogeneous part of the chemical potential of the neutral ion-electron pairs, respectively
\begin{equation}
    g_h = g_{p, chem} + g_{n, chem}, \ \mu_h = \pd{g_h}{c}.
\end{equation}
Both $g_h$ and $\mu_h$ are independent on $\phi$, since the total electrostatic energy of species ``p", ``n", ``d" is zero due to the assumed bulk electroneutrality. $\mu_h$ is also called "diffusional chemical potential" of the homogeneous mixture \cite{Bazant2017ThermodynamicElectro-autocatalysis}, defined as the change in free energy upon adding a neutral ion-electron pair at constant temperature and pressure. 

Now we consider a lumped form of \autoref{eq:G} at equilibrium state. We assume the concentration inside each phase is uniform, and treat the phase boundary as infinitely thin interface with energy $\gamma_i$. We denote the two phases as $r$ and $p$ and assume $c_r>c_p$. As before, we use $s$ to denote the walls as a third phase. Then for a 2D system, we have 
\begin{equation}
\begin{aligned}
     G_{lumped}  & = g_{h}(c^r) S^r + g_{h}(c^p)[S-S^r] \\
     &+ \gamma_i  A^{r,p} + \left[\gamma_{s}^0(c^r)- \gamma_{s}^0(c^p) \right] A^{s, r} \\
     &+ \lambda \left[c^r S^r + c^p[S-S^r] - c_m S \right] + const,
     \label{eq:G_lumped}
\end{aligned}
\end{equation}
where $A^{r,p}$ is the contact length between the rich and poor phases, $A^{s,r}$ is the contact length between the rich phase and the solid wall, $S$ is the total area and $S^r$ is the area of the rich phase, $c_m$ is the mean ion concentration in the nanofilm, and $\lambda$ is the Lagrange multiplier for species conservation. We then further consider the phase distribution as a rich phase spherical drop sitting on an electrode and surrounded by the poor phase (shown in \autoref{fig:contact_angle}), and get $A^{r,p} = 2r\theta$, $A^{s,r} = 2r\sin\theta$, $S^r = r^2\theta - r^2\sin\theta\cos\theta$, where the curvature radius $r$ and contact angle $\theta$ are both defined in the rich phase. 
%We further assume no overpotential so that $\gamma_{s,r} = \gamma_s^0(c_r)$ and  $\gamma_{s,p} = \gamma_s^0(c_p)$. 
Therefore, now $G_{lumped}$ is a function of five variables: $c^r, c^p, \lambda, r, \theta$.

Next, by minimizing $G_{lumped}$ (let first derivatives equal zero), we get the following relation for the binodal points $c_{b0}$ and $c_{b1}$ (assume $c_{b0}<c_{b1}$) 
\begin{equation}
     \mu_h(c_{b0}) = \mu_h(c_{b1}) = \frac{g_{b1}-g_{b0} + \frac{\gamma_i}{r_0}}{c_{b1}-c_{b0}}=\mu_b,
     \label{eq:binodal}
\end{equation}
and the Young-Laplace equation
\begin{equation}
    \gamma_{s}^0(c_{b1}) - \gamma_s^0(c_{b0}) = - \gamma_i \cos \theta_s^0,
    \label{eq:Young_Laplace}
\end{equation}
where $\theta_s^0$ is the equilibrium contact angle defined in the rich phase and $r_0$ is the equilibrium curvature of the rich phase. Here we have used $\pd{\gamma_s^0}{c}|_{c_{b0},c_{b1}}=0$, which is justified in \autoref{eq:grad_c_intf} and \autoref{eq:theta_s0}. These two equations are also true for a system of a poor-phase droplet surrounded by rich phase, but note that in this case $r_0$ is negative. The effect of the curvature of the rich phase on binodal points (term $\frac{\gamma_i}{r_0}$ in \autoref{eq:binodal}) is essentially analogous to the effect of the curvature of a liquid droplet on the vapor pressure (Kelvin equation or Ostwald–Freundlich equation at constant temperature \cite{Thomson1872OnLiquid, Pinson2015HysteresisPaste, Pinson2018InferringMedia}, and Gibbs-Thomson equation at constant pressure \cite{Mitchell2008NuclearCryoporometry}). For the simulations in this paper, we can find that the contribution of surface energy and interface curvature to phase diagram are small (\autoref{sec:numerical}). In this case, we can drop the surface energy related terms in  \autoref{eq:binodal}, and get the same relations as the flat interfaces \cite{Bazant2017ThermodynamicElectro-autocatalysis}.

We can also get the spinodal points $c_{s0}$, $c_{s1}$ (assume $c_{s0}<c_{s1}$) by letting the second derivatives of $G_{lumped}$ to be zero. For flat interfaces, we have \cite{Bazant2017ThermodynamicElectro-autocatalysis}
\begin{equation}
   \left. \pd{\mu_h}{c} \right|_{c_{s0}}  = \left. \pd{\mu_h}{c}\right|_{c_{s1}} = 0.
   \label{eq:spinodal}
\end{equation}
%\begin{equation}
%   \left. \left(\pd{\mu_h}{c} + \frac{\partial^2 \gamma_s^0}{\partial c^2}\frac{A^{s,p}}{S^r}\right) \right|_{c_{s0}}  = \left. \left(\pd{\mu_h}{c}  + \frac{\partial^2 \gamma_s^0}{\partial c^2}\frac{A^{s,r}}{S^r}\right)\right|_{c_{s1}} = 0.
%   \label{eq:spinodal}
%\end{equation}

\subsection{Switching current}
In Ref.\cite{Tian2022InterfacialNanofilms}, we have derived an  analytical estimation of the switching current above which  MP occurs at least on one electrode, based on the 1D model for LTO memristors. This estimation is  consistent with simulations. Here, we generalize the estimation for any forms of $\mu_{k,chem}(c)$ and $D_k(c)$. Again, the estimation only works for 1D phase distribution (flat phase boundaries) and neutral wetting surfaces.

We first analyze the steady state concentration profiles in each phase below the switching current (see Figure S4 in Ref.\cite{Tian2022InterfacialNanofilms}). Here we neglect the defect concentration ($c_d = 0$, $c_p = c_n = c$). At steady state, the ion diffusion and ion migration balance each other. Therefore, inside each phase, the current can be expressed as
\begin{equation}
    i =  \frac{D_n c F}{RT} \pd{\mu_{h}}{c} \pd{c}{x} = \pd{\mathcal{F}(c)}{x},
\end{equation}
where $\mathcal{F}(c) = \int  \frac{D_n c F}{RT} \pd{\mu_h}{c} dc $. This also directly indicates that current can induce CP inside each phase. 

If CP is stronger enough to trigger phase separation  on either electrode (the concentration reaches nearby spinodal point), MP occurs. We further assume that the concentration near the phase boundaries is still at binodal points. 
Therefore, the switching current can be estimated by
\begin{equation}
    i_c^{ana} =  \min\left\{\frac{\mathcal{F}|_{c_{b0}}^{c_{s0} }}{\mathcal{O}(c_m)},  \frac{\mathcal{F}|_{c_{s1}}^{c_{b1} }}{L - \mathcal{O}(c_m)}  \right\},  
    \label{eq:ic_ana}
\end{equation}
where  $\mathcal{O}(c_m)= \frac{c_{b1}-c_m}{c_{b1}-c_{b0}} L$  is the estimation of the spatial occupation of the ion-poor phase 0 when the average of $c$ in the system is $c_m$. 

%If $D_k(c)$ and non-ideal part of $\mu_{chem,k}(k)$ are polynomials of $c$,  $\mathcal{F}(c)$ also has a simple form, as we can see for the regular solution model in \autoref{sec:regular_solution}.

\subsection{Interfacial energy and thickness}{\label{sec:interface}}
In this part, we derive the energy density and estimate the thickness of the phase boundaries. The analysis is based on flat phase boundaries, but should still work for curved phase boundaries if the curvature radius is much larger than the interface thickness.

First, we obtain the concentration gradient perpendicular to a flat phase boundary (along coordinate $x_I$) in equilibrium \cite{Cahn1958FreeEnergy}:
\begin{equation}
    \kappa_p \pd{c}{x_I} = \sqrt{2\kappa_p [g_h(c) - g_h(c_{b0}) - (c-c_{b0})\mu_{b}]} = \mathcal{I}(c),
    \label{eq:grad_c_intf}
\end{equation}
by integrating 
\begin{equation}
    \mu_b \nabla c = (\mu_h -  \kappa_p \nabla^2 c)\nabla c
\end{equation}
along the coordinate $x_I$.

Then we calculate the interfacial energy of the phase boundary: 
\begin{equation}
    \gamma_{i} = \int_{-\infty}^{\infty} \left[ g(c)  - g_h(c_{b0}) - \mu_b(c - c_{b0}) \right] dx_I,
\end{equation}
which can lead to \cite{Cahn1958FreeEnergy}
\begin{equation}
    \gamma_{i} = \int_{c_{b0}}^{c_{b1}} \mathcal{I}(c) dc.
\end{equation}

Using function $\mathcal{I}(c)$, we can also estimate the thickness of the phase boundary $h_I$ from \autoref{eq:grad_c_intf}, and get $h_I \approx \kappa_p c_0/\mathcal{I}(c_0/2)$, where $c_0$ is the characteristic concentration (maximum ion concentration) and $c_0/2$ is approximately the concentration at the middle of the interface. Since the concentration gradient is largest at the middle of the interface, this expression should slightly underestimate $h_I$. In \autoref{sec:results}, we compare this estimation with simulations and find good agreement. %the scaling analysis of the regular solution model \cite{Bazant2013TheoryThermodynamics}.

\subsection{Contact angle}{\label{sec:contact_angle}}
\begin{figure}
    \centering
    \includegraphics[width = 0.5 \columnwidth]{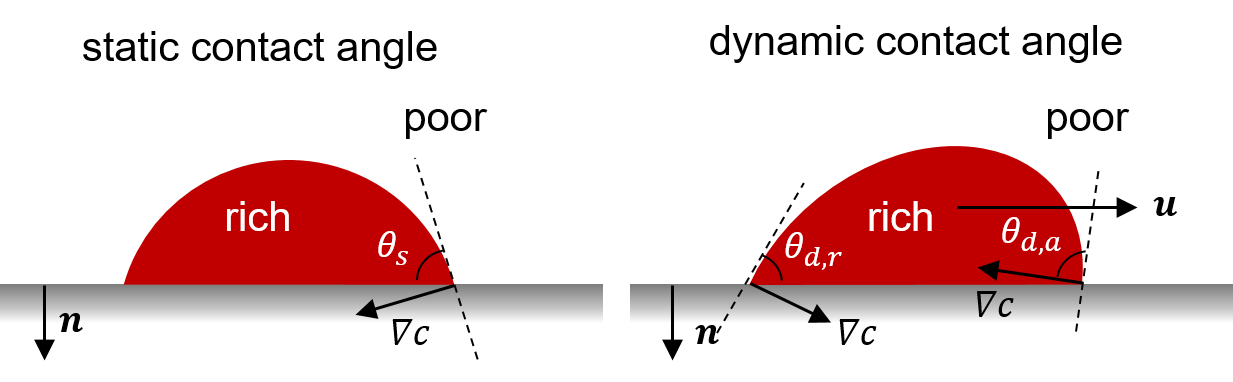}
    \caption{Schematic for the static and dynamic contact angle, where the contact angle is defined from $\cos \theta  = \mathbf{n} \cdot \nabla c/|\nabla c| $. }
    \label{fig:contact_angle}
\end{figure}

In this part, we intend to analyze the effects of surface charge ($C\neq 0$, $\eta\neq 0$) and surface energy relaxation ($\tau_\gamma>0$) on MP, by analyzing the contact angle defined in the rich phase based on the gradient of $c$, as shown in \autoref{fig:contact_angle}:
\begin{equation}
    \frac{\mathbf{n} \cdot \nabla c}{|\nabla c|} = \cos \theta.
    \label{eq:def_theta}
\end{equation}
Then we consider three contact angles: $\theta_s^0$, the static contact angle without overpotential; $\theta_s$, the static contact angle at overpotential $\eta$; $\theta_d$, the dynamic contact angle. In the derivation, we assume that the concentration gradient perpendicular to the phase boundaries (\autoref{eq:grad_c_intf}) is not perturbed from the equilibrium state in any case. And we assume $\pd{C}{c} = 0$ here for simplicity. 
%We see $\theta_s^0$ as a given parameter, and then analyze the behaviors of $\theta_s$ and $\theta_d$. 

Without overpotential and surface relaxation ($\tau_\gamma = 0$ and $\eta = 0$),  \autoref{eq:dgamma_eff_dc} becomes
\begin{equation}
    \pd{\gamma_s^0}{c} = - \mathcal{I}(c) \cos \theta_s^{0}.  
    \label{eq:theta_s0}
\end{equation}
Therefore, one additional parameter $\theta_s^0$ is enough to give the wetting boundary condition. Obviously,  the integral of this equation leads to the classical Young-Laplace equation (\autoref{eq:Young_Laplace}).

Next, we add an overpotential to the nanofilm-electrode interface ($\tau_\gamma = 0$ but $\eta \neq 0$). After plugging \autoref{eq:theta_s0}  in \autoref{eq:dgamma_eff_dc}, we obtain
\begin{equation}
    \mathcal{I}(c)(\cos \theta_s - \cos \theta_s^0) + \frac{C \eta}{F} \pd{\mu_{n, chem}}{c} = 0.
    \label{eq:theta_s}
\end{equation}
Therefore, if we assume $\pd{\mu_{n, chem}}{c}>0$ (the Fermi energy lifts due to ion intercalation)  and $C\neq 0$, we can deduce that: $\eta>0$ leads to $\theta_s>\theta_s^0$, which means that the positive electrode (the current goes from this electrode to the nanofilm) is less wetting to the rich phase due to the applied currents; $\eta<0$ leads to $\theta_s<\theta_s^0$, which means that the negative electrode (the current goes from the nanofilm to this electrode) is more wetting to the rich phase due to the applied currents.

Finally, we analyze the dynamic contact angle ($\tau_\gamma > 0$ but $\eta \neq 0$). We plug \autoref{eq:theta_s} into \autoref{eq:dgamma_eff_dc} and get
\begin{equation}
    \mathcal{I}(c) (\cos \theta_d - \cos \theta_s) = -\tau_\gamma \pd{c}{t}.
\end{equation}
Therefore,  $\theta_d<\theta_s$ leads to $\pd{c}{t}<0$, which means that the contact line moves to the rich phase (receding contact angle); $\theta_d>\theta_s$ leads to $\pd{c}{t}>0$, which means that the contact line moves to the poor phase (advancing contact angle).

\subsection{Discussion on electrowetting}

In the previous part, we derive \autoref{eq:theta_s} for the effect of surface charge and electric currents on  static contact angle for our model system shown in \autoref{fig:system}. In this part, we want to compare this effect to the well-known phenomenon of electrowetting for liquid droplets, which can also help to understand our problem. 

In the classical setup for electrowetting, a droplet of aqueous salt solution is placed on a dielectric substrate, and a voltage is applied between an electrode wire in the droplet and the substrate. Then, people found that the voltage, regardless of sign, can make the dielectric more wetting to the droplet \cite{Mugele2005Electrowetting:Applications, Sedev2011Electrowetting:Dynamics, Adamson1976PhysicalSurfaces}. This phenomenon has been widely applied in microfluidics and ``lab-on-a-chip" devices. 

Apparently, this sign-independence of the voltage-tuned contact angle in electrowetting is not seen in the system of this study (according to \autoref{eq:theta_s}),  though the droplet in electrowetting is analogous to our more-conductive phase droplet (can be ion-rich or ion-poor phase depending on the material). The reason is that the two systems have  different boundary conditions. Our system uses parallel electrodes, while electrowetting uses an electrode wire in the droplet and a substrate electrode. Therefore,  in our system, the overpotential should be almost uniform on each electrode (since the nanofilm has small variation electrochemical potential if surface resistance dominates total resistance),  while in electrowetting, the potential drop only occurs at the droplet-substrate interface. 

If we modify our system by removing the top electrode and putting an electrode wire with no surface resistance in the rich phase droplet, assume that the rich phase is more conductive, neglect the charge at the interface between the poor phase and electrode, and consider the equilibrium state without leaky current (electron transfer at interfaces), then we get 
\begin{equation}
    \mathcal{G} = G_{lumped} -  C\eta ^2 A^{s,r},
\end{equation}
where $G_{lumped}$ is modified from \autoref{eq:G_lumped} by replacing $\gamma_s^0(c^r)$ by $\gamma_s(c^r)=\gamma_s^0(c^r) + \frac{1}{2}C\eta^2$ to include the electrostatic energy, and $C\eta^2 A^{s,r}$ is the energy released from the voltage source. Then, the effective surface energy at the droplet-substrate surface now becomes $\gamma_{s,r}^{\mathrm{eff}} = \gamma_{s}^0(c^r) - \frac{1}{2} C\eta^2$, which leads to the famous equation for electrowetting \cite{Mugele2005Electrowetting:Applications, Huang2012AControl}
\begin{equation}
    \cos{\theta_s} = \cos \theta_s^0 + \frac{C\eta ^2}{2\gamma_i}.
\end{equation}
%which can be directly applied as a boundary condition to the phase field model for electrowetting  \cite{Huang2012AControl}. 
Note that the $\eta$ that we define is the same as $U-U_{pzc}$ in \cite{Mugele2005Electrowetting:Applications} ($U$: applied voltage, same as $\Delta \phi$ in this work; $U_{pzc}$: applied voltage at point of zero charge, same as $(\Delta \phi)_0$ in this work).  Therefore, our phase field model is consistent with the electrowetting theory if the same boundary conditions and simplifications are applied. 

In addition to boundary conditions, our system is also very different from  the electrowetting system in the following aspects. First, our system is a pure solid system while electrowetting manipulates liquid droplets. Therefore, the phase field in our system can be represented by scaled ion concentration, while the phase field in electrowetting should be represented by scaled water concentration. Furthermore, the drag on the liquid droplet by electric fields should be weaker than that on the solid rich phase (ions only occupy a small fraction in liquid), so usually  electrowetting neglects the body force on the droplet due to bulk electric fields. Finally, the leaky current at the nanofilm-electrode interface is very important in our system because it can lead to phase redistribution in the bulk, while in electrowetting the leaky current is usually avoided. 
%but has additional effects such as electric-field-driven phase redistribution. 

\section{Nondimensionalization and regular solution model}{\label{sec:eqs}}
\subsection{Dimensionless governing equations for the general phase-field model}
In this part, we summarize all the governing equations and boundary conditions in the dimensionless form. We define the scales of the variables as below: the concentration scale $c_0 = c_p^{max}$,  the length scale $L$ as the thickness of the nanofilm, the time scale as the diffusion time of ions $\tau_D = L^2/D_p^0$, the electrochemical potential scale $RT$ ($R$ is the gas constant and $T$ is the temperature), the electric potential scale $V_T = RT/F$ (thermal voltage), the free energy density scale $R T c_0$, the gradient energy coefficient scale $\kappa_0 = L^2 R T/c_0$, the interface energy density scale $\gamma_0 = R T c_0 L$, the capacitance density scale $C_0 =  F c_0 L/V_T$, the species flux scale $J_{k,0}=D_k^0 c_0/L$ ($D_k^0$ are constants chosen based on the material), the current scale $i_D = J_{n, 0}F$, and the energy scale $e_0 = i_D V_T \tau_D$. Then we get the dimensionless variables by dividing the dimensional variables by the corresponding scales. Specifically, we also define $\tilde{\mathcal{F}}(\tilde{c}) = \mathcal{F}(c) L/i_D$, $\tilde{\mathcal{O}}(\tilde{c}_m) = \mathcal{O}(c_m)/L$ to non-dimensionalize \autoref{eq:ic_ana_F}.

Now we can list the dimensionless governing equations:
\begin{equation}
    \tilde{c}_p = \tilde{c}_n - \tilde{c}_d = \tilde{c}.
    \label{eq:nd_e_neutral}
\end{equation}
\begin{equation}
    r_k \frac{\partial \tilde{c}_k}{\partial \tilde{t}} + \tilde{\nabla} \cdot \tilde{\mathbf{J}}_k = 0, \ \ \ \tilde{\mathbf{J}}_k = - \tilde{D}_k \tilde{c}_k \tilde{\nabla} \tilde{\mu}_k, \ \ \ (k = p,n)
    \label{eq:nd_NP}
\end{equation}
where  $\tilde{\nabla} = L \nabla$, $r_p = 1$, $r_n= D_p^0/D_n^0$. And $\tilde{\mu}_k$ is the scaled electrochemical potential
\begin{equation}
    \tilde{\mu}_k = \tilde{\mu}_{k,chem} + z_k \tilde{\phi} - \tilde{\kappa}_k \tilde{\nabla}^2 \tilde{c},
\end{equation}
where we assume $\tilde{\kappa}_n = 0$.

Next we list the boundary conditions. We put the electrodes on coordinates $\tilde{y}=0$ and $\tilde{y} = 1$, where
\begin{subequations}
\begin{equation}
    \tilde{\mathbf J}_p \cdot \mathbf{n} = 0,
\end{equation}
\begin{equation}
    \tilde{\mathbf J}_n \cdot \mathbf{n} = -\mathbf{n} \cdot \tilde{\mathbf{i}} = \tilde{J}_n^{ET} (\tilde{c}, \tilde{\eta}) + \tilde{C}\pd{\Delta \tilde{\phi}}{\tilde{t}}, 
\end{equation}
\begin{equation}
    \tilde{\kappa}_p \tilde{\nabla} \tilde{c} \cdot \mathbf{n} = -\tilde{\tau}_\gamma \pd{\tilde c}{\tilde t}   +\tilde{\mathcal{I}}(\tilde{c})\cos \theta_s^0 -\tilde{C}\tilde{\eta} \pd{\tilde{\mu}_{n, chem}}{\tilde c},
\end{equation}
\end{subequations}
where $\tilde{\mathcal I}(\tilde c)$ has the following expression
\begin{equation}
    \tilde{\mathcal I}(\tilde c) = \sqrt{2\tilde{\kappa}_p [\tilde{g}_h(\tilde{c}) - \tilde{g}_h(\tilde{c}_{b0}) - (\tilde{c}-\tilde{c}_{b0}) \tilde{\mu}_{b}]},
    \label{eq:static_contact}
\end{equation}
and $\tilde{J}_n^{ET}(\tilde{c}, \tilde{\eta})$ is given in \autoref{sec:regular_solution}. Finally, we close the domain with two symmetric boundaries on $\tilde{x}=0$ and $\tilde{x}= W/L$ where
\begin{equation}
    \tilde{\mathbf J}_n \cdot \mathbf{n}= 0,\ \tilde{\mathbf J}_p \cdot \mathbf{n} = 0,\ \tilde{\nabla} \tilde{c} \cdot \mathbf{n} = 0.
\end{equation}

\subsection{Constitutive equations for a regular solution}{\label{sec:regular_solution}}
Up to now, our phase-field model is general and can be applied to any form of $D_p(\tilde{c})$, $D_n(\tilde{c})$, $\mu_{p,chem}(\tilde{c})$, $\mu_{n,chem}(\tilde{c})$, and $J_n^{ET}(\tilde{c}, \tilde{\eta})$. Now we want to specify the above functions for a regular solution model and use them for simulations in the next section. In addition, we also derive $\tilde{\mathcal{F}}(\tilde{c})$ for the regular solution model to get the switching current. 

In the regular solution model, we consider both electrons and ions as Fermi-Dirac particles hopping on specific lattices and experiencing some interaction energy. This model for ions (or ion-electron pairs) has been widely used \cite{Bazant2013TheoryThermodynamics}. For electrons, this model should also work well for many ion-intercalation materials where mobile electrons are usually localized on transition metal atoms and the electronic conduction is mainly due to electron hopping \cite{Cox2010TransitionProperties, Moulson2003Electroceramics:Applications, Zhou2006ConfigurationalLixFePO4, Tsai2019AbBatteries} but not band conduction \cite{Sze2007PhysicsDevices}.  Therefore, we can assume \cite{Bazant2013TheoryThermodynamics, Tian2022InterfacialNanofilms}
\begin{equation}
    \tilde{D}_k = 1 - \frac{\tilde{c}}{\tilde{c}_k^{max}}
\end{equation}
and
\begin{equation}
    \tilde{\mu}_k = \ln \frac{\tilde{c}_k}{\tilde{c}_k^{max} - \tilde{c}_k} + \tilde{\mu}_k^0  + \Omega_k (\tilde{c} + \rho \tilde{c}(1-\tilde{c})) ,
\end{equation}
where $\tilde{c}_k^{max}$ is the maximum available sites for ions or electrons to sit in, $\Omega_k$ controls the magnitude of interaction energy, and $\rho$ helps control the symmetry of the phase diagram. We further define $\Omega = \Omega_p + \Omega_n$, which is a key parameter to determine $\mu_h$. $\tilde{c}_k^{max}$ may be different for ions and electrons.  For example, we assume $c_n^{max}/ c_p^{max}= 5/3$ in our LTO model in \cite{Tian2022InterfacialNanofilms}, according to the available space for ions to intercalate in and the Ti atoms for electrons to hop on. In this case, the ion-rich phase is more electronically conductive. 

%In addition, note that we assume the electron conduction is due to electron hopping\cite{Cox2010TransitionProperties, Maier2004PhysicalSolids}, which is common in transition metal oxides and different from the band conduction in traditional semiconductors \cite{Sze2007PhysicsDevices}. 

Since we assume the mobile electrons are localized, the electron transfer can be seen as Faraday reaction (e.g., $\mathrm{Ti}^{4+} + e^- (\mathrm{Pt}) \rightleftharpoons \mathrm{Ti}^{3+}$ for LTO memristors). We then apply the generalized, empirical Butler-Volmer equation for electron transfer kinetics (see details in \cite{Tian2022InterfacialNanofilms})
\begin{subequations}
\begin{equation}
   \tilde{J}_n^{ET}(\tilde{c}, \tilde{\eta}) = \mathrm{Da} f(\tilde{c}; \alpha) g(\tilde{\eta}_{\mathrm{eff}} ; \alpha) ,
\end{equation}
\begin{equation}
    f(\tilde{c}; \alpha) =  (\tilde{c}_n/\tilde{c}_n^{max})^\alpha (1 - \tilde{c}_n/\tilde{c}_n^{max})^{1-\alpha} e^{\alpha \Omega_n \left( \tilde{c} + \rho \tilde{c}(1-\tilde{c}) \right)  },
\end{equation}
\begin{equation}
    g(\tilde{\eta}_\mathrm{eff}; \alpha) = e^{(1-\alpha) \tilde{\eta}_\mathrm{eff}} - e^{-\alpha \tilde{\eta}_\mathrm{eff}},
\end{equation}
\label{eq:gBV}
\end{subequations}
where the effective overpotential $\tilde{\eta}_\mathrm{eff}$ only works on the electron transfer dynamics, excluding the voltage loss on the series resistance or film resistance \cite{Doyle1996ComparisonCells}
\begin{equation}
    \tilde{\eta}_\mathrm{eff} = \tilde{\eta} - \tilde{J}_n^{ET} \tilde{R}_s.
\end{equation}
Therefore, if $\Omega_n>0$, which indicates that the Fermi energy increases as more ions are intercalated, the electron transfer rate increases almost exponentially with $c$. 

Finally, we calculate $\tilde{\mathcal{F}}(\tilde{c})$ based on the regular solution model, and get
\begin{equation}
\begin{aligned}
    \mathcal{F} &=  \left(1+\frac{\tilde{c}_p^{\max }}{\tilde{c}_n^{\max }}\right) \tilde{c}-\tilde{c}_p^{\max }\left(1-\frac{\tilde{c}_p^{\max }}{\tilde{c}_n^{\max }}\right) \ln \left(1-\frac{\tilde{c}}{\tilde{c}_p^{\max }}\right) \\
    & + \Omega (1+\rho) \left( \frac{\tilde{c}^2}{2} - \frac{\tilde{c}^3}{3\tilde{c}_n^{\mathrm{max}}}\right) - 2\Omega \rho \left(\frac{\tilde{c}^3}{3} - \frac{\tilde{c}^4}{4\tilde{c}_n^{\mathrm{max}}}\right),
\end{aligned}
\label{eq:ic_ana_F}
\end{equation}
which can be plugged in the non-dimensional form of \autoref{eq:ic_ana} to calculate the switching current $\tilde{i}_c^{ana}$.

\section{Numerical simulation}{\label{sec:results}}

\subsection{Simulation method and setup}{\label{sec:numerical}}
\subsubsection{Numerical method}
In this part, we perform simulations on the equations summarized in \autoref{sec:eqs}. We discretize the equations using finite volume method, which ensures species conservation, and use the convex splitting method to get an unconditionally stable time marching scheme \cite{Eyre1998UnconditionallyEquation, Yuille2003TheProcedure}.  Note that we do convex splitting not only for bulk energy but also for surface energy \cite{Gao2012AProblem}. We also use adaptive time stepping to save the computational time without losing the accuracy \cite{Zhang2013AnModel, Li2017ComputationallyEquation}. Finally, we implement the method in MATLAB R2022a. We use  the automatic differentiation package developed by Gorce \cite{Gorce2022SparseGitHub} to calculate the Jacobian.

\subsubsection{Parameters}{\label{sec:para}}
For all the simulation cases presented in this paper, we assume the nanofilm has thickness $L=\SI{50}{nm}$, and let $\tilde{\kappa}_p = 0.0014$ so that the phase boundary thickness  $h_I \approx \sqrt{\tilde{\kappa}_p/\tilde{\mathcal I}(0.5)}L\approx\SI{2.1}{nm}$. This estimation is much more accurate than the scaling analysis $h_I \approx \sqrt{\tilde{\kappa}_p/|\Omega|}L = \SI{0.54}{nm}$\cite{Bazant2013TheoryThermodynamics}. The simulation domain is $\tilde{x}\in [0, 2]$ (the direction parallel to the electrodes, $W/L=2$) and $\tilde{y} \in [0, 1]$ (the direction perpendicular to the electrodes). We put 200 volumes in the $x$ direction and 100 volumes in the $y$ direction, so that there are at least 5 points in the phase boundary. 

Though our theory is general, we need to parameterize $\tilde{\mu}_k$ and $\tilde{J}_n^{ET}(\tilde{c}, \eta)$ to run simulations. We use the regular solution model and choose the following parameters based on the LTO material, as we did in  the previous paper \cite{Tian2022InterfacialNanofilms}:   $\tilde{c}_p^{max}=1$, $\tilde{c}_n^{max}=5/3$, $\Omega_n = 20$, $\Omega_p = -32$, $D_p = \SI{1e-16}{m^2\ s^{-1}}$, $D_n = \SI{1e-11}{m^2\ s^{-1}}$, $c_0 = \SI{22.8}{M}$, $\tilde{c}_{d}=0.01$, $\tilde{R}_s = 100$. In addition, we choose $\rho = 0.4$ to adjust the symmetry of the phase diagram. For the electron transfer, we assume symmetric transfer $\alpha^T = \alpha^B = 0.5$, and choose $\mathrm{Da}^T = \SI{1e-2}{}$ and $\mathrm{Da}^B = \SI{1e-4}{}$ unless specified. Finally, when we want to consider dynamic contact angle, we set $\tilde{\tau}_{\gamma} = 0.02$; when we want to consider surface charge, we take $C = 10 \epsilon_0/\SI{1}{nm} = \SI{0.0885}{C\ V^{-1}\ m^{-2}}$ so that $\tilde{C} = \SI{2.07e-5}{}$.

Based on the above parameters, we can derive the following properties of the material. (1) The dimensional scales are $\tau_D = \SI{25}{s}$, $J_p ^0  = \SI{4.56e-5}{mol\ s^{-1} m^{-2}}$, $J_n ^0  = \SI{4.56}{mol\ s^{-1}\ m^{-2}}$, $i_D = \SI{4.4e5}{A\ m^{-2}}$, $\gamma_0 = \SI{2.83}{N\ m^{-1}}$. The surface energy of the phase boundaries is $\tilde{\gamma}_{i} = 0.023$. (2) For phase equilibrium, we have $\tilde{c}_{b0}=0.0265$, $\tilde{c}_{b1} = 0.9733$, $\tilde{c}_{s0} = 0.1422$, and $\tilde{c}_{s1} = 0.8135$ for flat phase boundaries. Then the perturbation of the interface curvature to the binodal points and spinodal points is within 10\% for droplets with curvature radius bigger than \SI{10}{nm}. For example,  $r_0/L = 0.314$ leads to $\tilde{c}_{b0}=0.0280$, $\tilde{c}_{b1} = 0.9754$, and $r_0/L = -0.314$ leads to $\tilde{c}_{b0}=0.0250$, $\tilde{c}_{b1} = 0.9708$. The error between these predictions and simulations is with in $0.05\%$. (3) Next, we can analyze the transport behaviors. The ionic conductivity scales with $\tilde{c}(1-\tilde{c})$, and the ionic tracer diffusivity $D_p = D_p^0(1-\tilde{c})$ decreases with $c$. Since $\tilde{c}_n^{max}=5/3$ and the electronic conductivity scales with $\tilde{c}(1-\tilde{c}/\tilde{c}_n^{max})$, the material experiences insulator-metal transition as concentration increases. In addition, since $\Omega_n>0$, the electron transfer between the ion-rich phase and electrodes feels less resistance, compared with that between the ion-poor phase and the electrodes. (4)  Finally, curves of $\tilde{\mu}_h(\tilde{c})$, $\tilde{\mathcal{I}}(\tilde{c})$ and $\tilde{i}_c^{ana}(\tilde{c})$ for these given parameters can be found in Figure S1 (some other parameters are also plotted for a comparison). We estimate the switching current for  flat phase boundaries and neutral wetting condition, and get $\tilde{i}_c^{ana}(0.1) = 0.0975$, $\tilde{i}_c^{ana}(0.9) = 0.5310 $.

\subsubsection{Simulation plan}
In this paper, we consider two  two mean concentrations $\tilde{c}_m = 0.1, 0.9$. Both $\tilde{c}_m$ enable phase separation since they are between the binodal points.  %Since we have assumed $\tilde{c}_n^{max}>\tilde{c}_p^{max}$ and $\Omega_n>0$, the ion-rich phase should be more electronically conductive and the local electron transfer between the ion-rich phase and electrodes feels less resistance, compared with ion-poor phase.  
In systems with $\tilde{c}_m = 0.1$ ($\tilde{c}_m = 0.9$), there is a thin film or small droplet of the more-conductive ion-rich phase (less-conductive ion-poor phase) surrounded by the less-conductive ion-poor phase (more-conductive ion-rich phase). 

For each $\tilde{c}_m$, we consider two processes. We first simulate the response of the system to a step current (set total current and assume each electrode has uniform potential), to present the time evolution of phase boundaries and obtain the switching current, time, and energy.  As shown in \autoref{sec:analysis}, 1D MP is controlled by currents but not total voltage drop. In the current response, the electron transfer kinetics do not matter at all for 1D MP without surface charge \cite{Tian2022InterfacialNanofilms} (electron transfer kinetics is critical for MP-induced RS but not MP itself). Ion-modulated electron transfer can influence the current density distribution on electrodes for 2D phase distribution, and can also influence the overpotential magnitude and thus influence the surface energy if surface charge exists.       Then we simulate the process of cyclic voltammetry and show the non-volatile RS behaviors. In this process, electron transfer kinetics is important for all the cases.  In the step current simulation, we first let the system relax for dimensionless time of 5, then apply a constant current for $0\leq \tilde{t}<5$, and finally remove the current and let the system relax for $5 \leq \tilde{t} < 25$. In the cyclic voltammetry simulation, we first let the system relax for dimensionless time of 2, and then run three cycles at voltage sweeping rate $2 V_t/s$. Then we only plot results for the last cycle. 

\begin{figure*}[hbt!]
    \centering
    \includegraphics[width = 1 \textwidth]{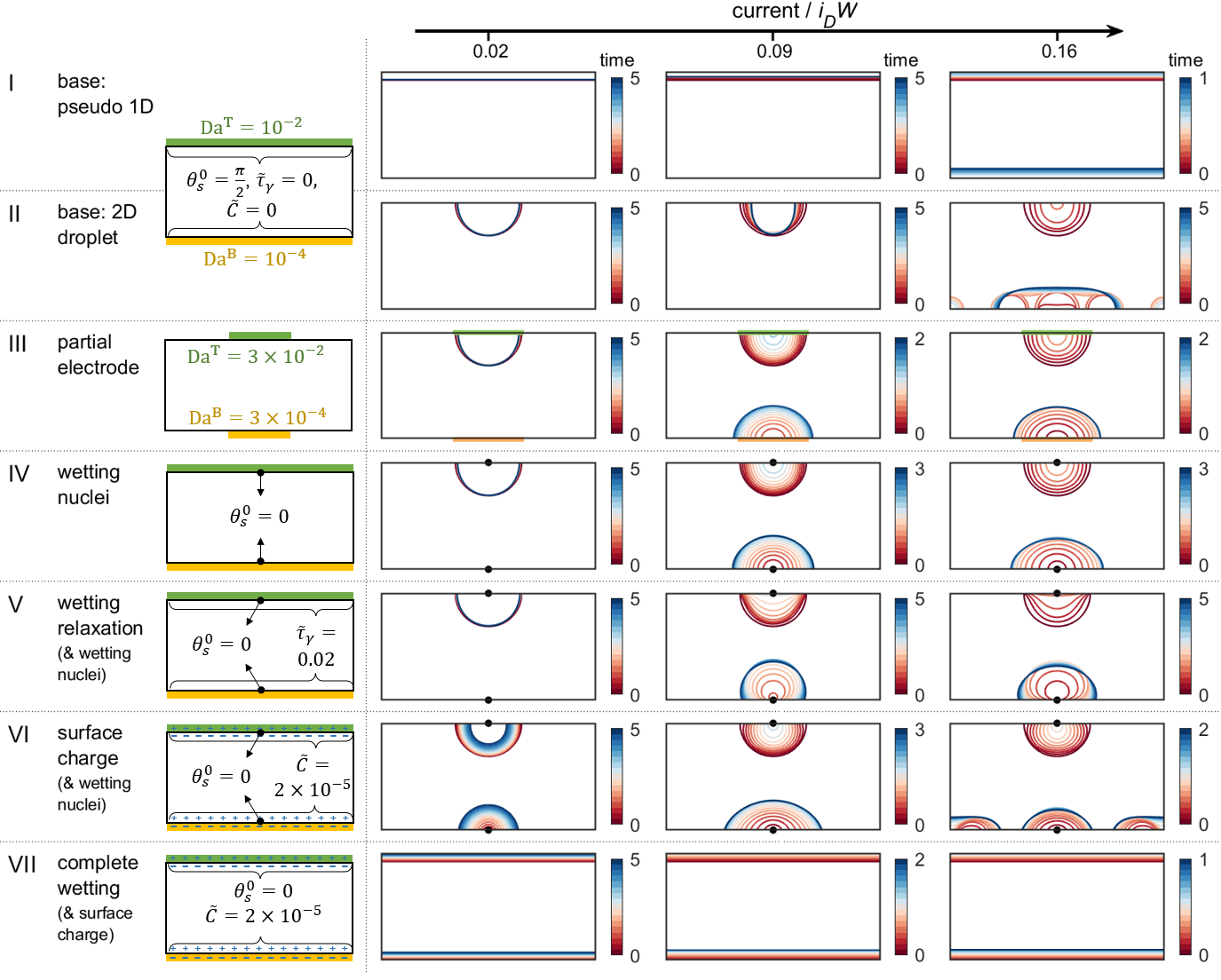}
    \caption{Time ($\tilde{t} = t/\tau_D$) evolution of the phase boundaries (contour lines of $\tilde{c}=0.5$) in the  ion-poor systems (mean concentration $\tilde{c}_m = 0.1$) after applying constant electric currents ($\tilde{I} =I/i_D W $, where $I = \int_0^W i \mathrm{d}x$), for three currents and seven cases. A sufficiently large downward current can move the ion-rich film or droplet from the top electrode to the bottom.   
    %and Case I is a base case where the initial phase boundaries are flat and each electrode is homogeneous with neutral wetting ($\theta_s^0 = \pi/2$), no wetting relaxation ($\tilde{\tau}_\gamma = 0$), and no surface charge ($\tilde{C} = 0$). %Case I can be seen as 2D simulations of the 1D problems in Ref.\cite{Tian2022InterfacialNanofilms}.
    Case I and II has the same boundary conditions but different initial phase boundaries. Case I can be seen as 2D simulations of the 1D problems in Ref.\cite{Tian2022InterfacialNanofilms}. Case III-VII have one or two boundary conditions different from I and II, as described in the left column. The corresponding animations of concentration and electric potential profiles can be found in Movie S1 and Movie S2. 
    %case III has partial electrodes with 1/3 original length and tripled Da (pinked parts are not conductive), case IV, V, and VII have nuclei wetting to the rich phase (location marked by the green dots), case V has wetting relaxation $\tilde{\tau}_\gamma = 0.02$, case VI and VII have surface charge capacitance $\tilde{C} = \SI{2.07e-5}{}$, and case VII has  electrodes completely wetting to the rich phase ($\theta_s^0 = 0$). The dimensional scales (values for a LTO film with thickness $L = \SI{50}{nm}$) are: $\tau_D = L^2/D_p$ (25 s), $i_0 = D_n^0 F c_0/L$ ($\SI{0.44}{\mu A\ \mu m^{-2}}$). }%More details for the setup of the simulations can be found in \autoref{sec:numerical}.  $\theta_s^0 = \frac{\pi}{2}(1 -  e^{2((y-1)/0.05)^2 })$  
    }
    \label{fig:contour_c0.1}
\end{figure*}

Finally, for each $\tilde{c}_m$ and each process, we consider seven cases with six sets of boundary conditions and two initial conditions. We first choose a base case with flat initial phase boundaries, uniform surface resistance, neutral wetting, no surface charge and no surface relaxation. The base case can be seen as a 2D simulation of the 1D problem in Ref.\cite{Tian2022InterfacialNanofilms}. Then we modify the base case by implementing curved initial phase boundaries, heterogeneous surface resistance, heterogeneous wetting condition (wetting nuclei),  surface energy relaxation,  surface charge, and complete wetting condition (completely wetting to one phase), and get six other cases. Therefore we can analyze each effect separately. Note that we add the complete wetting case, because solid interfaces are likely to show complete wetting instead of finite contact angle \cite{Cogswell2013TheoryNanoparticles}. The parameters for the seven cases are listed in \autoref{fig:contour_c0.1}, \autoref{fig:contour_c0.9}. When adding wetting nuclei, we use  $\theta_s^0 = \frac{\pi}{2}(1 \pm  e^{-2((\tilde{x}-1)/0.05)^2 })$, where ``+" (``$-$") is used to make the nuclei completely wetting to the poor (rich) phase. 

%In addition, we consider case VII with complete wetting because it is possible that the real electrode-nanofilm boundary shows complete wetting to one phase instead of neutral wetting \cite{Cogswell2012CoherencyNanoparticles}. Therefore, we also add case VII to study complete wetting conditions.

%For the boundary conditions, we first choose a base case with uniform boundary conditions, where we assume neutral wetting ($\theta_s^0 = \pi/2$), no surface relaxation ($\tilde{\tau}_\gamma = 0$), and no surface charge ($\tilde{C} = 0$).  Then we add heterogeneity of ET rates (tripled $\mathrm{Da}^T$, $\mathrm{Da}^B$ for $\frac{2}{3}\leq \tilde{y}\leq \frac{4}{3}$ and zero $\mathrm{Da}^T$, $\mathrm{Da}^B$ elsewhere), wetting nuclei (non-uniform $\theta_s^0$ with $\theta_s^0 = \frac{\pi}{2}(1 \pm  e^{-2((y-1)/0.05)^2 })$), surface energy relaxation ($\tilde{\tau}_\gamma = 0.02$), surface charge ($\tilde{C}= \SI{2.07e-5}{}$), or complete  surface wetting ($\theta_s^0 = 0$ or $\pi$), and get five more systems of different boundary conditions. Therefore, we can analyze each physics separately. For the base system, we consider two initial conditions, one with flat phase boundary (a pseudo-1D problem solving by 2D simulation) and the other one with 2D droplet of rich or poor phase. For the complete wetting case, we also give the flat boundaries initially. For all the other systems, we give the 2D droplet initial conditions. 

\begin{figure*}[hbt!]
    \centering
    \includegraphics[width = 0.9\textwidth]{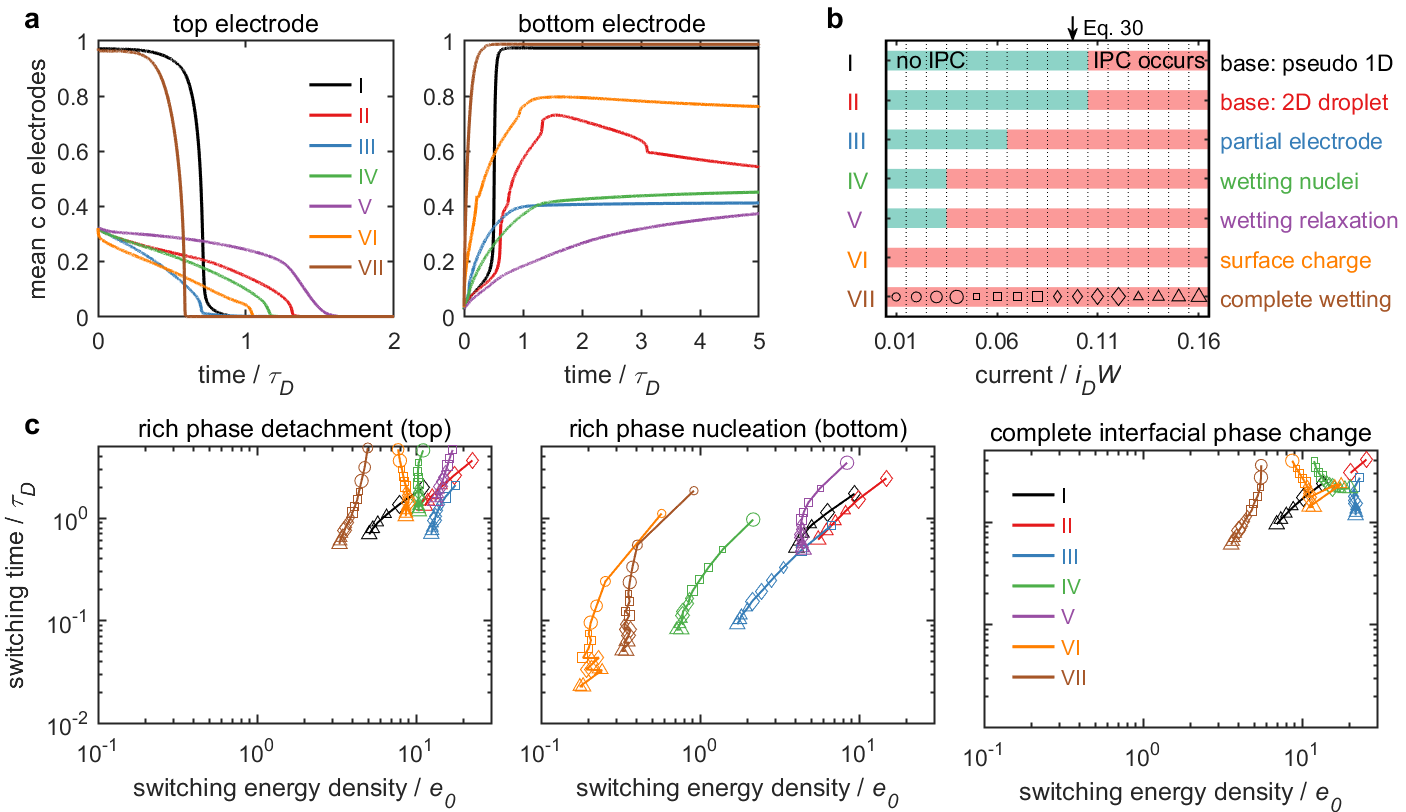}
    \caption{Characterization of the MP-induced interfacial phase change (IPC) in the ion-poor systems shown in \autoref{fig:contour_c0.1}.  (a) Time ($\tilde{t} = t/\tau_D$) evolution of the mean ion concentration on electrodes at current $\tilde{I}= I/i_D W  = 0.16$. (b) Switching current ($\tilde{I}_s$) obtained by checking whether IPC occurs at least on one electrode in the simulations. The arrow shows the analytical switching current for case I from \autoref{eq:ic_ana}. (c) Switching time ($\tilde t_s$) along with energy density ($\tilde{e} =  e /e_0 = \int_0^{\tilde{t}_s}\tilde{I} \tilde{V} \mathrm{d}\tilde{t}$) characterized by three events:  the completion of the detachment of the rich phase from the top electrode,  the initiation of the nucleation of the rich phase on the bottom electrode, and the completion of IPC on both electrodes. The markers in (c) represent different currents, which are consistent with those shown at the bottom of (b). The dimensional scales (values for a 50-nm-thick LTO film) are: $\tau_D = L^2/D_p$ (25 s), $i_D = D_n^0 F c_0/L$ ($\SI{0.44}{\mu A\ \mu m^{-2}}$), $e_0 = i_D V_T  \tau_D$ ($\SI{0.275 }{\mu J\ \mu m^{-2}}$).  }
    \label{fig:checkboard_c0.1}
\end{figure*}

\subsection{Simulation results for the ion-poor system}
In this part, we present results for the  the ion-poor system  ($\tilde{c}_m = 0.1$). We place the rich phase near the top electrode initially, and expect that the rich phase can be moved to the bottom at high enough electric currents.

\subsubsection{Time evolution of phase boundaries and interfacial concentration}
%In this part, we show the dynamics of phase change in response to a step current.  We show the time evolution of the phase boundary defined by the contour line of $\tilde{c} = 0.5$ in \autoref{fig:contour_c0.1} ($\tilde{c}_m=0.1$) and \autoref{fig:contour_c0.9} ($\tilde{c}_m=0.9$) , and that of the mean concentration on the top and bottom electrodes in \autoref{fig:c_step_0.1} ($\tilde{c}_m=0.1$)  and \autoref{fig:c_step_0.9} ($\tilde{c}_m=0.9$). 

\autoref{fig:contour_c0.1} shows the time evolution of phase boundaries  in response to a step current. We only present the time periods after the application of current and before the completion of IPC on both electrodes (see \autoref{fig:checkboard_c0.1} and \autoref{sec:switching_cur_time_c0.1} for detailed time analysis). 

First, we can see that only high enough currents can trigger phase redistribution (MP). Then we compare different cases. For case I, II with the same homogeneous boundary conditions, as the current is applied, the flat initial phase boundaries maintain flat (case I), while  the curved initial phase boundaries maintain curved (case II). Without surface heterogeneity, the rich phase nucleation can occur at random, multiple locations (case II, current = 0.16).  The heterogeneous electron transfer resistance (partial electrode, case III) and heterogeneous wetting condition (wetting nuclei, case IV) can make phase nucleation easier to occur and preferable to occur at certain locations (centers in case III, IV). Also note that in case II, III, IV, the contact angle is always 90 degrees on electrodes where we apply the static neutral wetting condition ($\nabla c \cdot \mathbf{n} = 0$,  except in the nuclei in IV). As we add the wetting relaxation in case V, we get the advancing contact angle ($\theta_d > \pi/2$) on the bottom electrode, and the receding angle ($\theta_d < \pi/2$) on the top electrode. As we add a surface charge capacity in case VI (but no wetting relaxation), we find that the top electrode with positive overpotential becomes less wetting to the rich phase, which supports the detachment of the rich-phase droplet, while the bottom electrode with negative overpotential becomes more wetting to the rich phase, which supports the nucleation of the rich-phase (may occur at multiple locations). This is consistent with our analysis for \autoref{eq:theta_s}, and indicates that the surface charge makes MP easier to occur if $\pd{\mu_{n,chem}}{c}>0$. Finally, we find that the completely wetting surface (wetting to the rich phase) also makes MP easier to occur (case VII), similar to the effects of the wetting nuclei (case VI).  

%\subsubsection{Time evolution of interfacial concentrations}
Then we use \autoref{fig:checkboard_c0.1}(a) to quantitatively compare the mean interfacial concentration along with time for different cases at current 0.16. As we can see, different cases show very different curve patterns. For example, the pseudo-1D phase distribution in case I and case VII leads to the most abrupt IPC (steepest slope during switching), while the wetting relaxation in case V leads to slowest IPC. 
%In addition, we find that in case VI, the interfacial concentration changes fastest at the beginning of the applied current. 
Basically, this model opens the opportunity to potentially learn the physics of MP using only the information for interfacial concentration evolution.

We also plot time evolution of interfacial concentration and applied voltage at different currents for each case in Figure S2 and Figure S3, where the relaxation period after the removal of current is also included. Figure S2 shows clearly that  IPC only occurs if  the concentration near at least one electrode goes through a spinodal point. Also, we can see that the IPC is non-volatile as long as MP completes  during the application of the current, and the increased current accelerates the process.

\subsubsection{Switching current, time, and energy}{\label{sec:switching_cur_time_c0.1}}
In this part, we fill in more simulations for series of currents, and obtain the switching current, time, and energy. 
We identify three critical events for MP-induced IPC: (1) the completion of the detachment of the rich phase film/droplet from the top electrode judged from $\max (\tilde{c}|_{\tilde{x}=1}) < 0.5$, (2) the start of the  nucleation of the rich phase on the bottom judged from $\max (\tilde{c}|_{\tilde{x}=0}) > 0.5$, and (3) the completion of IPC judged from the slope of the time evolution of mean interfacial concentration $\sqrt{\left[(\partial \langle \tilde{c}|_{x=0} \rangle /\partial \tilde{t})^2  +(\partial \langle \tilde{c}|_{x=1} \rangle/\partial \tilde{t})^2 \right]/2 } \leq 0.01$).  If at least one of the first two events occur, we say that IPC and MP occurs at the given current, and call the corresponding threshold current as the switching current.  Then we define three switching times as the time spent respectively for the three events to occur during the application of the current. Note that the switching time has definition only if the corresponding event occurs. We show the switching current for different cases in \autoref{fig:checkboard_c0.1}(b), and show the switching time versus energy in \autoref{fig:checkboard_c0.1}(c).  

First, we can see from \autoref{fig:checkboard_c0.1}(b) that the switching current indicated by simulations for the pseudo-1D base system is well consistent with our analytical prediction (\autoref{eq:ic_ana}), which has been well studied in our previous work \cite{Tian2022InterfacialNanofilms}. Then we can analyze  the differences between the 1D and 2D pictures. By comparing case I with II, and case IV with V, we find the initial condition and wetting relaxation do not matter for the switching current. Then by comparing case II with III and IV, and case IV with VI and VII, we find that the heterogeneity of the boundary conditions, the surface charge, and the complete wetting condition can reduce the switching current significantly. 

Next, we discuss the switching time for each case shown in \autoref{fig:checkboard_c0.1}(c). Generally, the switching time is dominated by ion diffusion and can be reduced by increasing the current. Regardless of the existence of the wetting nuclei, the nucleation of the rich phase on the bottom occurs earlier in time than the detachment of the rich phase from the top at the given current. Complete IPC and rich phase detachment from the top take similar time, and the complete wetting condition (case VII) takes smallest time among all the cases for complete IPC and rich phase detachment. The increasing current can reduce the switching time by nearly one magnitude for rich phase detachment and complete IPC, and nearly two magnitudes for rich phase nucleation. 

Then we discuss the switching energy for each case shown in \autoref{fig:checkboard_c0.1}(c).  The switching energy density is mainly determined by the ion diffusion time multiplied by the electron diffusion current. Since the inverse of switching time should be roughly proportional to the excess current (current minus switching current) \cite{Tian2022InterfacialNanofilms}, the switching energy  may decrease with current at small currents, and should not change much with current at large currents. Sometimes the switching energy can also increase with current if multiple nucleation points start to appear.    The switching energy is much smaller for the rich phase nucleation compared with rich phase detachment and complete IPC. By increasing current, the switching energy does not change much with current for rich phase detachment and complete IPC, but can decrease significantly for rich phase nucleation. We can also compare the dimensional numbers with experiments. If we assume the switching energy density is $e_0$, we predict that the switching energy for a $\SI{100}{\mu m}\times \SI{100}{\mu m}$ device is around $\SI{1}{mJ}$, while the experimental value is around $\SI{0.1}{mJ}$ ($(\SI{4}{V})^2\times \SI{0.8}{\mu J}\times 50 \times \SI{500}{ms}$, for 50 pulses in Figure 2(c) of Ref.\cite{Gonzalez-Rosillo2020Lithium-BatterySeparation}). These two values are pretty close, considering that the electronic conductivity varies greatly in solid devices (e.g., see Ref.\cite{Young2013ElectronicAnode}).

Furthermore, we compare different cases for switching time and energy in \autoref{fig:checkboard_c0.1}. We find that partial electrode (case III),  wetting nuclei (case IV),  surface charge (case VI), and complete wetting (case VII) can reduce the time and energy for rich phase nucleation very significantly compared with the base case (case I), while the wetting relaxation (case V) can slow down the switching. 

Finally, the above analysis for the three events of IPC is limited to $\tilde{c}_m=0.1$. We can imagine that if we further decrease the mean concentration (still above $c_{b0}$), the time and energy for complete IPC and rich-phase detachment should decrease and become closer to those for rich-phase nucleation. 

\begin{figure*}[bt!]
    \centering
    \includegraphics[width = 0.98 \textwidth]{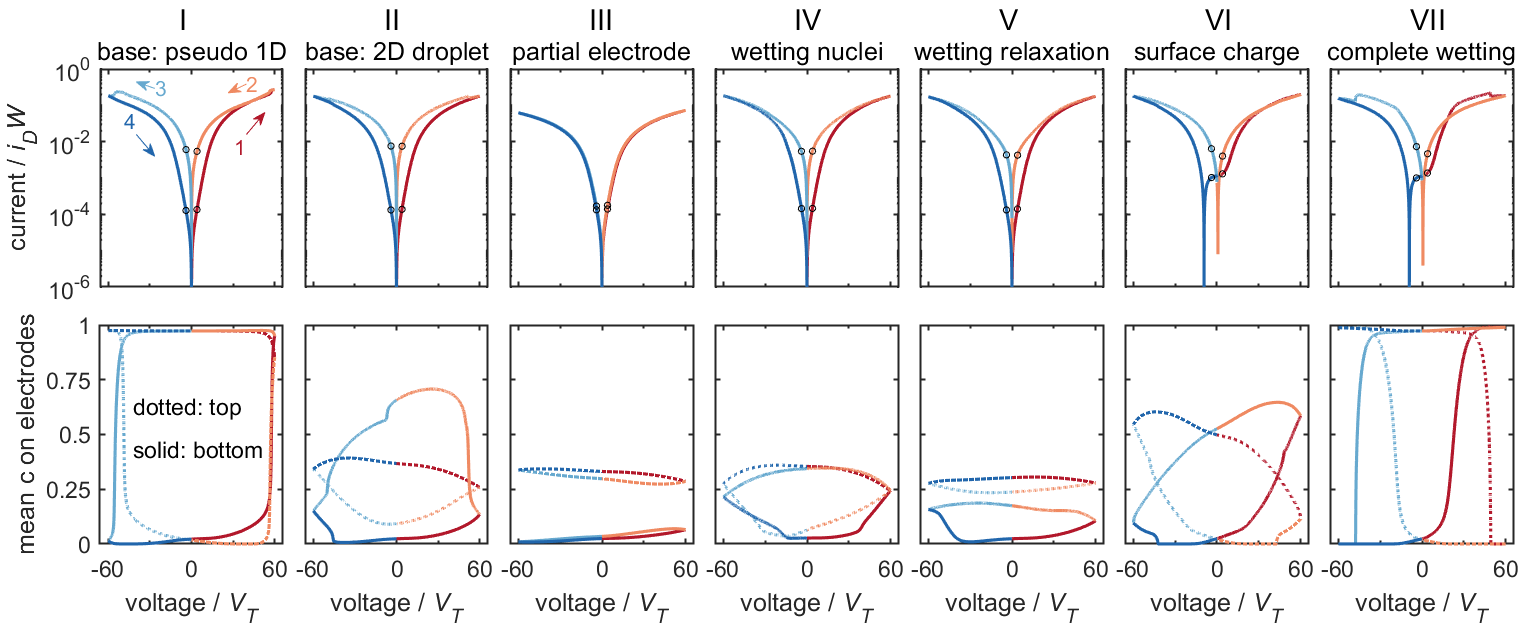}
    \caption{Current ($\tilde{I} = I/i_D W$, top row) and mean concentration on electrodes (bottom row) along with the applied voltage ($\tilde{V} = V/V_T$, where $V_T = 0.025 V$) during cyclic voltammetry with voltage sweeping rate 50 (scaled by $V_T/\tau_D$), for the ion-poor systems shown in \autoref{fig:contour_c0.1}. The circles on the current-voltage curves mark the points with $|\tilde{V}|=4$, which guide the eyes to read the resistance ratio.  The dimensional scales (values for a 50-nm-thick LTO film) are: $\tau_D = L^2/D_p$ (25 s), $i_D = D_n^0 F c_0/L$ ($\SI{0.44}{\mu A\ \mu m^{-2}}$). The corresponding time evolution of concentration contour maps can be found in Movie S3. }
    \label{fig:CV_c0.1}
\end{figure*}

\subsubsection{Resistive switching}
Finally, we discuss on the memristive switching behaviors of the ion-poor system due to MP-induced IPC. The simulation results for cyclic voltammetry with a maximum voltage magnitude of 60 is shown in \autoref{fig:CV_c0.1}. Since the bottom electrode is assumed to have smaller Da, it should has smaller electron transfer rate given the same $\tilde{c}$ and $\tilde{\eta}$. In addition, the electron transfer rate also increases with $\tilde{c}$ since we assume positive $\Omega_n$. Therefore, we expect that the migration of the rich phase droplet/film from the top electrode to the bottom should reduce the overall resistance. In another word, the nonvolatile IPC can lead to nonvolatile resistive switching. 

The results shown in \autoref{fig:CV_c0.1} is consistent with the above expectation. For all the cases, no matter the rich phase exists as a droplet or a film, RS occurs as long as IPC occurs. The current-voltage curve in case I is consistent with our 1D simulations in the previous paper  \cite{Tian2022InterfacialNanofilms}. When surface charge exists (case VI, VII), the current-voltage curves show hysteretic behaviors (zero current occurs at non-zero voltage).  In addition, we can see that the wetting nuclei (case IV), surface charge (case VI), and complete wetting condition (case VII) can make IPC occur earlier (at smaller voltage).

%The asymmetry of the curve comes from the overlimiting current in the poor phase at high voltage. In this condition, the current exceeds the diffusion limiting current in the poor phase (around $ \tilde{c}_{b0}/\mathcal{O}(\tilde{c}_{b0})\approx 0.03$),  and the fixed charge $\tilde{c}_d$ dominates the total bulk resistance. We also track the interfacial concentration and find IPC finishes right before the applied voltage reaches 60. By comparing 

%Then we notice that  Then we find that the 2D cases b, c, 

\begin{figure*}[bt!]
    \centering
    \includegraphics[width =  \textwidth]{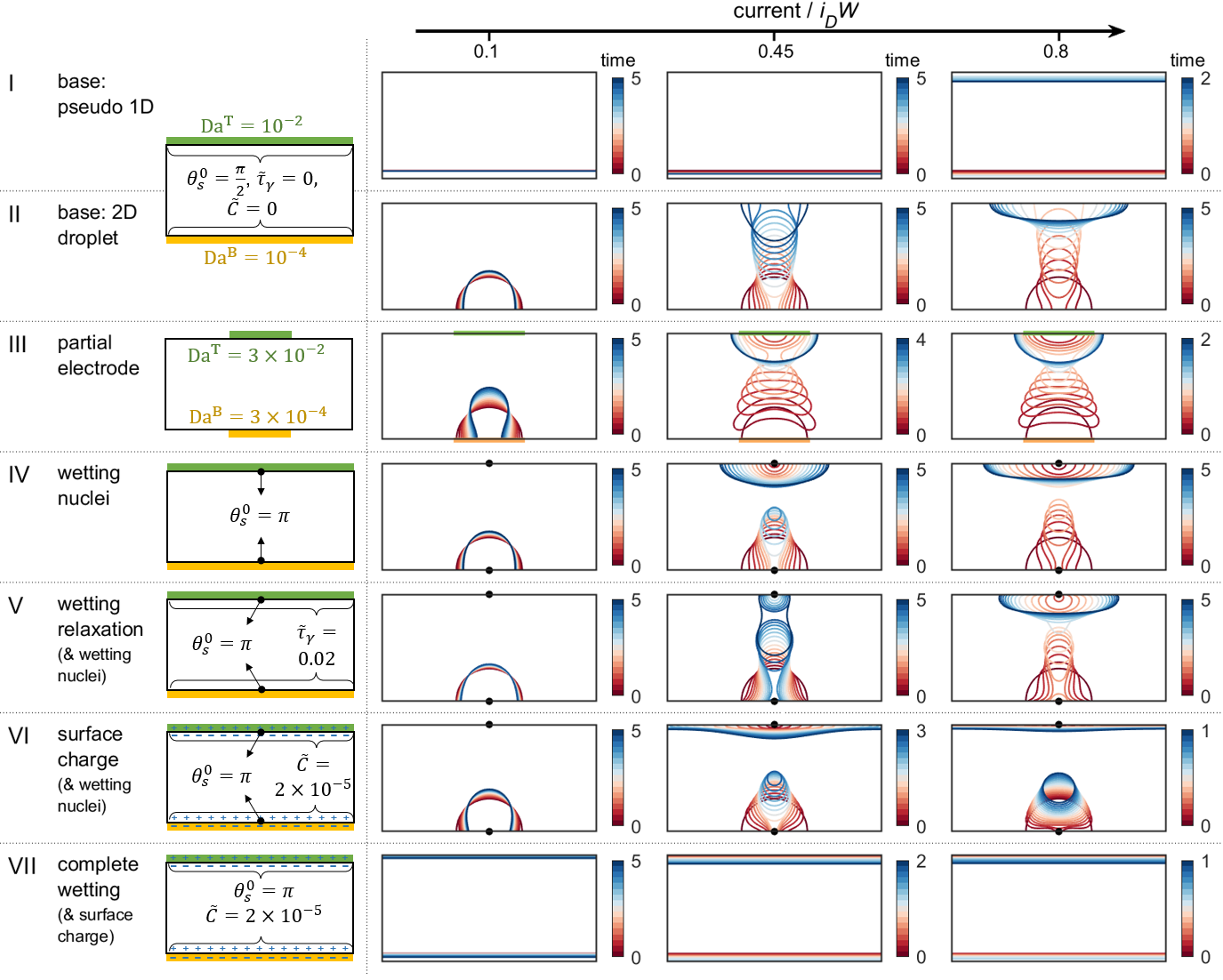}
    \caption{Time ($\tilde{t} = t/\tau_D$) evolution of the phase boundaries (contour lines of $\tilde{c}=0.5$) in the ion-rich systems (mean concentration $\tilde{c}_m = 0.9$) after applying constant currents ($\tilde{I} =I \rangle/i_D W$, where $I = \int_0^W i\mathrm{d}x$), for three currents and seven cases.  A sufficiently large downward current can move the poor phase film or droplet from the bottom electrode to the top. 
     Case I and II has the same boundary conditions but different initial phase boundaries. Case I can be seen as 2D simulations of the 1D problems in Ref.\cite{Tian2022InterfacialNanofilms}. Case III-VII have one or two boundary conditions different from I and II, as described in the left column. The corresponding animations of concentration and electric potential profiles can be found in Movie S4 and Movie S5.   }
    \label{fig:contour_c0.9}
\end{figure*}

\begin{figure*}[hbt!]
    \centering
    \includegraphics[width = 0.9\textwidth]{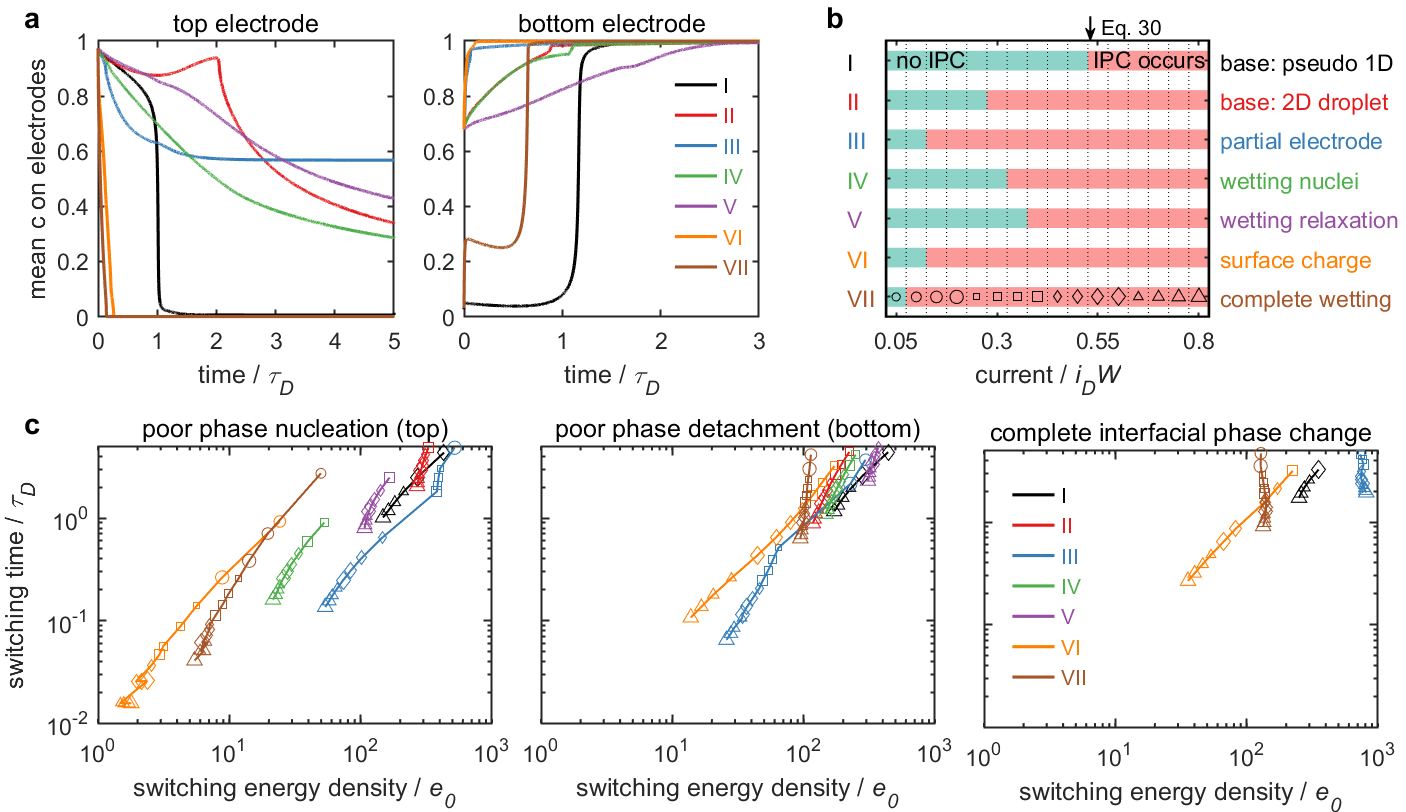}
    \caption{Characterization of the MP-induced interfacial phase change (IPC) in the ion-rich systems shown in \autoref{fig:contour_c0.9}.  (a) Time ($\tilde{t} = t/\tau_D$) evolution of the mean ion concentration on electrodes at current $\tilde{I}= I /i_D W = 0.8$. (b) Switching current ($\tilde{I}_s$) obtained by checking whether IPC occurs at least on one electrode in the simulations. The arrow shows the analytical switching current for case I from \autoref{eq:ic_ana}. (c) Switching time ($\tilde t_s$) along with energy density ($\tilde{e} =  e /e_0 = \int_0^{\tilde{t}_s}\tilde{I} \tilde{V} \mathrm{d}\tilde{t}$) characterized by three events:  the initiation of the nucleation of the poor phase on the top electrode,  the completion of the detachment of the poor phase from the bottom electrode, and the completion of IPC on both electrodes. The markers in (c) represent different currents, which are consistent with those shown at the bottom of (b). The dimensional scales (values for a 50-nm-thick LTO film) are: $\tau_D = L^2/D_p$ (25 s), $i_D = D_n^0 F c_0/L$ ($\SI{0.44}{\mu A\ \mu m^{-2}}$), $e_0 = i_D V_T  \tau_D$ ($\SI{0.275 }{\mu J\ \mu m^{-2}}$). }
    \label{fig:checkboard_c0.9}
\end{figure*}
\begin{figure*}[hbt!]
    \centering
    \includegraphics[width = 0.98 \textwidth]{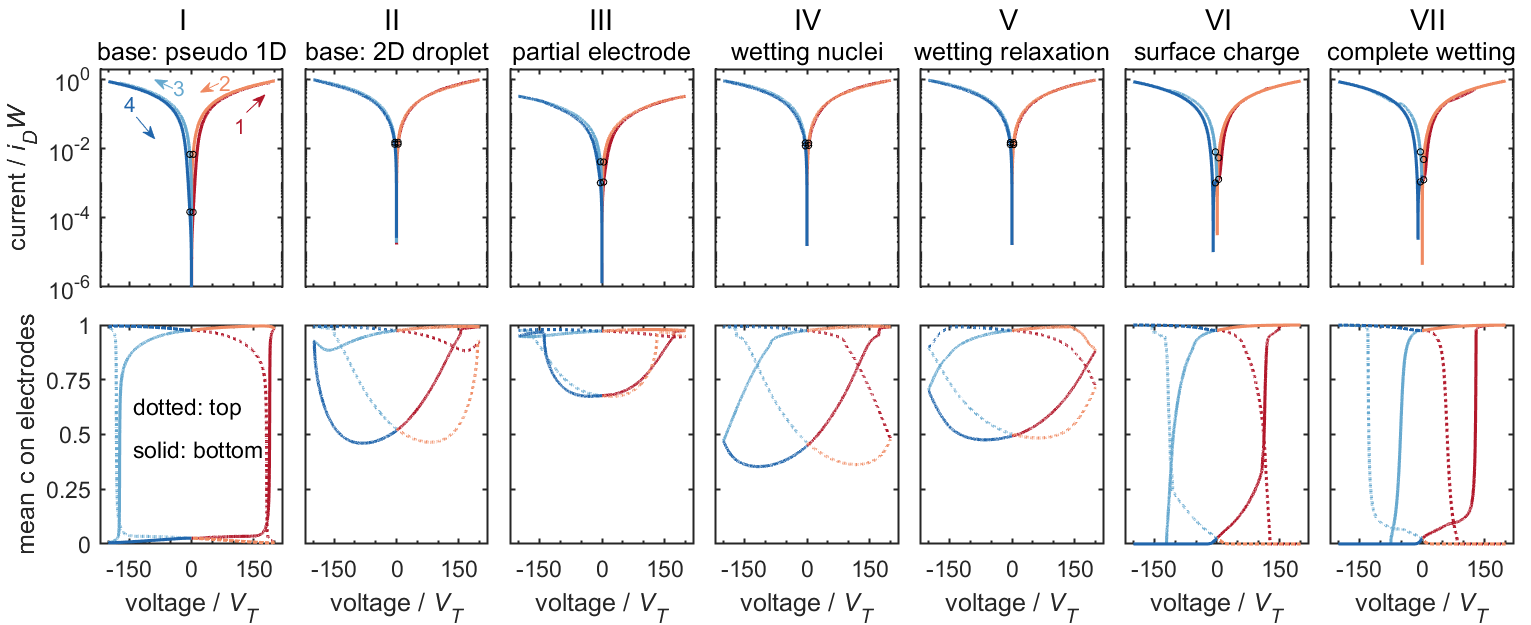}
    \caption{Current ($\tilde{I} = I/i_D W$, top row) and mean concentration on electrodes (bottom row) along with the applied voltage ($\tilde{V} = V/V_T$, where $V_T = 0.025 V$) during cyclic voltammetry with voltage sweeping rate 50 (scaled by $V_T/\tau_D$), for the ion-rich systems shown in \autoref{fig:contour_c0.9}. The circles on the current-voltage curves mark the points with $|\tilde{V}|=4$, which guide the eyes to read the resistance ratio.  The dimensional scales (values for a 50-nm-thick LTO film) are: $\tau_D = L^2/D_p$ (25 s), $i_D = D_n^0 F c_0/L$ ($\SI{0.44}{\mu A\ \mu m^{-2}}$).  The corresponding time evolution of concentration contour maps can be found in Movie S6.}
    \label{fig:CV_c0.9}
\end{figure*}

\subsection{Simulation results for the ion-rich system}
In this part, we present results for the ion-rich system ($\tilde{c}_m$ = 0.9). We initially place the poor phase  on the bottom electrode, and expect that the poor phase can be moved to the top at high enough currents. Since the ion-poor system has been discussed before, here we mainly focus on the different behaviors between the ion-poor system and the ion-rich system. These differences should come from the asymmetry between ion transport and electron transport, as described in \autoref{sec:para}.

%which is common in transition metal oxides. For the LTO material that we present here, the ionic conductivity ($\sim \tilde{c}(1-\tilde{c})$) almost disappears for both $\tilde{c}=0$ and $\tilde{c} = 1$, and the electronic conductivity ($\sim \tilde{c}(1-0.6\tilde{c})$) is almost zero for $\tilde{c}=0$ (insulator) but large for $\tilde{c}=1$ (conductor).

\subsubsection{Time evolution of phase boundaries and interfacial concentration}
As we can see in \autoref{fig:contour_c0.9}, MP in the ion-rich systems needs larger current than the ion-poor systems, and the time evolution of phase boundaries is very different. When there is no wetting nuclei (case II, III), the poor-phase droplet first detaches completely from the bottom electrode, and then migrates to the top electrode. However, in case II and III for the ion-poor systems and case IV, V, VI for both ion-poor and ion-rich systems,  the droplet detachment and nucleation can occur at the same time. In addition,  we find that the surface charge in case VI also makes MP easier to occur in the ion-rich system. During the application of the current, the bottom electrodes become less wetting to the  ion-poor phase, which helps the detachment of the droplet; while the top electrode becomes more wetting to the ion-poor phase, which helps the nucleation of the droplet. In addition, we find that the large currents plus the surface charge can squeeze the droplet to a film on the top electrodes. Finally, the complete wetting surface (wetting to the poor phase) can also help MP to occur.

The time evolution of the mean concentration on electrodes for the ion-rich system shown in \autoref{fig:checkboard_c0.9}(a) and Figure S4 is similar to that for the ion-poor system. However, during the application of the current, the mean concentration on the top in case II, IV, V, VI can increase more significantly due to the squeezing of the droplets by the large current. The squeezed droplets can rebound after the current is removed if the droplets have not been squeezed into a film (case II, IV, V), making the relaxation require longer time  (Figure S4, Figure S5).

\subsubsection{Switching current, time, and energy}
We define the switching current and times for the ion-rich systems similar to what we do for the ion-poor systems. Here we judge 
the full detachment of the poor phase from the bottom  from $\min (\tilde{c}|_{\tilde{x}=0}) > 0.5$, and the start of the poor phase nucleation on the top from $\min (\tilde{c}|_{\tilde{x}=1}) < 0.5$. If one of the above two events occurs, we say that MP and IPC occurs. 

As shown in \autoref{fig:checkboard_c0.9}, MP in the ion-rich system needs about 5 times larger switching current than the ion-poor system, as predicted by both the theory and simulations. By comparing case I and II, we find that the 2D poor-phase droplet is easier to detach from the bottom and nucleate on the top compared with the poor-phase film. The surface heterogeneity in case III and IV, surface charge in case VI, and complete wetting in case VII can also reduce the switching current.

Then we discuss the switching time and energy shown in   \autoref{fig:checkboard_c0.9}(c). The poor-phase droplet detachment from the bottom can occur earlier in time than the poor-phase droplet nucleation on the top if the electrodes have neutral wetting condition (case II, III). Otherwise, nucleation is always earlier than detachment.  The significant squeezing of the droplets before the formation of films in case II, IV, V increases the time to to complete IPC (longer than the step current time). Once the droplet is squeezed into a film on the top (case VI), IPC takes much shorter time and energy. In case VI, after the completion of IPC, there may still exist the migration of the poor phase in the system, but the interfacial concentration does not change anymore, as shown in \autoref{fig:contour_c0.9}.

\subsubsection{Resistive switching}
Unlike the ion-poor system, a less-conductive poor-phase droplet in the ion-rich system may not change the  interfacial resistance on each electrode significantly. To make a big change, the poor phase needs to cover the whole electrode, or the very conductive points for electron transfer. Therefore, we only expect case I, III, VI, VII in \autoref{fig:contour_c0.9} to show RS due to MP-induced IPC at the largest current. This expectation is consistent with the cyclic voltammetry shown in \autoref{fig:CV_c0.9}. At even higher current and voltage, the droplet may cover the whole electrode in case II, IV, V and lead to RS, too. 

Since the ion-rich system needs much larger current and voltage for MP-induced RS, we expect that the ion-poor system is more useful for MP-based memories made from LTO-like materials.

\section{Discussion and conclusion}{\label{sec:discussion}}
In this section, we first discuss the insights brought by this work on the understanding and optimization of MP-based memristors, then discuss possible extensions and improvements of the current model for general problems of coupled ion-electron transport, and finally make a conclusion. 

\subsection{Understanding and optimization of  MP-based memristors}

To begin with, we want to point out the generalizability and limitations of the analysis of MP-based memristors in this work. The scaling analysis, including the scales for switching energy and time, should be general. However, our simulations are all based on LTO (Li$_{4+3x}$Ti$_5$O$_{12}$) material, which goes through insulator-metal transition for $x: 0\rightarrow 1$. We also assume that the surface resistance decreases with ion concentration, too. Therefore, the ion-poor system has lower electronic conductivity than the ion-rich system. Nevertheless, materials like Li$_x$CoO$_2$ with $x\in [0.5, 1]$ can show metal-insulator transition for increasing ion concentration, so the ion-rich system can have lower electronic conductivity. In addition, different materials may have different concentration-dependence for ion diffusivity and Fermi energy. Therefore, some of our conclusions obtained from simulations are only constrained to the LTO-like materials.  

First, we want to see some general conclusions regardless of the specific ion-intercalation materials. From scaling analysis, we know that the switching time is mainly limited by ion diffusion time, $\tau_D = L^2/D_p^0$, and can be reduced by one or two orders of magnitude by increasing the electric currents or including surface charge and heterogeneity (\autoref{fig:checkboard_c0.1}, \autoref{fig:checkboard_c0.9}). The switching energy per device area has scale $e_0 = i_D V_T \tau_D =  \frac{D_n^0 L F c_0 V_T}{D_p^0} $, which can also be reduced by manipulating surface conditions (\autoref{fig:checkboard_c0.1}, \autoref{fig:checkboard_c0.9}). Generally, we prefer materials with smaller electronic conductivity and larger ionic conductivity for MP, to reduce the switching time and energy simultaneously. As a comparison, Li-ion battery electrodes usually require the ion-intercalation material to have both large electronic conductivity and large ion diffusivity \cite{Nitta2015Li-ionFuture}. In addition, we can make the device thinner to reduce the switching time and energy, and scale down the device to reduce energy cost. 

\begin{figure}
    \centering
    \includegraphics[width = 0.45 \textwidth]{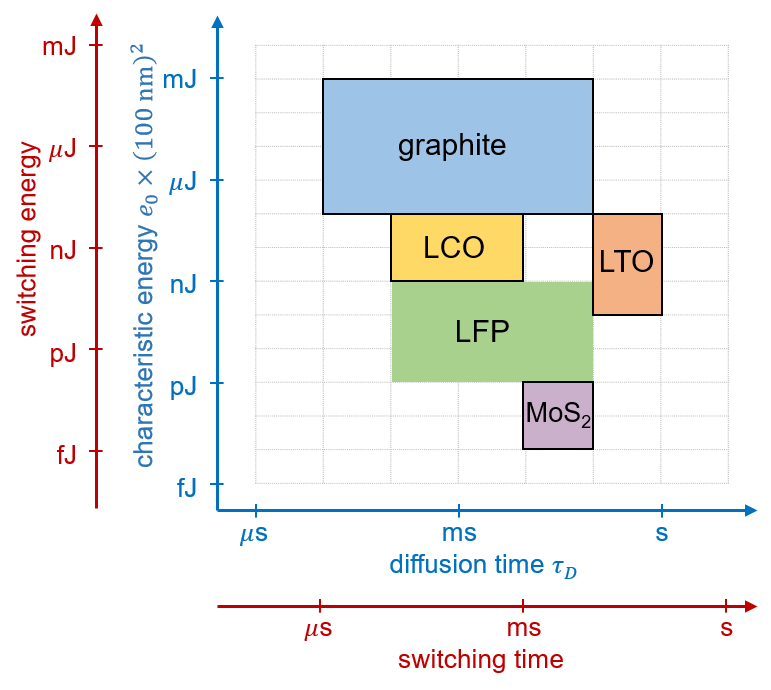}
    \caption{The estimation of the switching time and energy for multiphase polarization (MP), for several multiphase Li$^+$-intercalation materials shown in \autoref{tab:materials}. Among these materials, LTO, LCO, MoS$_2$, and graphite can be used to make MP-based memristors.}
    \label{fig:materials}
\end{figure}

We list values of $\tau_D$ and $e_0$ for 10-nm thick multiphase ion-intercalation nanofilms for different materials in \autoref{tab:materials}. Note that here the ``metallic" and ``insulating" phases are usually judged from electronic bands, and the metallic phase may also have conductivity much smaller than traditional metals due to small electron mobility (e.g., LTO and MoS$_2$). Though MP can occur in all those materials, LFP may not be used for MP-based memristors because it does not have significant resistance transition by changing ion concentration.  Then,  we use \autoref{fig:materials} to show the estimation of the switching time and energy density of MP for all the materials listed in \autoref{tab:materials}. Here, we estimate the switching time as $\tau_D/10$ and energy density as $e_0/10$, to include all the possible reduction enabled by adjusting boundary conditions and mean concentration. Among LTO, LCO, MoS$_2$, and graphite, graphite should show the smallest switching time in the scale of $\SI{1}{\mu s}$, and MoS$_2$ should show the smallest switching energy in the scale of fJ for a 100 nm$\times$100 nm$\times$10 nm device. There is still a lot of room to find better materials for MP-based memristors.

In the introduction, we have compared the principles for the MP mechanism and other nonvolatile RS mechanisms, as shown in \autoref{fig:mechanism}. We then want to compare the performance of these mechanisms. First, we note that the MP mechanism for a single crystal with homogeneous electrodes only has two resistance states. Therefore, multiple states for MP-based memristors should be obtained from polycrystalline structures or heterogeneous electrodes. This is similar to the FT mechanism. As a comparison, the RBT and PC memories can have numerous states by controlling the bulk ion concentration and conductive filament length, respectively. In addition, we note that the switching time for both MP and RBT are both limited by ion diffusion but can be improved by electric currents. Though ns-pulse can be used to switch the states of RBT, the reading can take longer time to wait for the stabilization of the system. Since ion diffusion in solids is usually slow, MP and RBT may have smaller switching time compared with PC or FT. However, they should require smaller energy, especially compared with PC. Ion-modulated phase change should cost less energy than crystallization or amorphorization. Finally, FT should have smaller switching time and energy than MP since local dipole rotation should be faster than ion migration. However, here we only consider the performances of switching energy and time, while other performances, material price and manufacturing process can make the comparison of the mechanisms more complex. 

\renewcommand{\arraystretch}{1.2}
\begin{table*}[hbt]
    \centering
    \caption{Scales of diffusion time $\tau_D $ and characteristic energy density $e_0 = i_D V_T/\tau_D$ (where  $V_T$ is thermal voltage and $i_D$ is electron diffusion current) for different ion-intercalation materials with ion diffusivity $D_p$, electronic conductivity $\sigma_n$, and thickness $L=\SI{10}{nm}$.  In the first column, ``i" represents the less conductive (insulating) phase, ``m" represents the more conductive (metallic) phase. Among the materials in this table, only LTO, LCO, MoS$_2$,and graphite can be used to make MP-based memristors. 
    For the model in this work, $\tau_D = L^2/D_p^0$ and $i_D = D_n^0 c_0 F/L$ are well defined. For other cases, we estimate  $\tau_D \approx L^2/D_p$, and $i_D \approx \sigma_n^* V_T/L$ where $\sigma_n^*$ is $\sigma_n$ for a more-conductive component $x$. 
    }
    \begin{tabular}{m{3.6cm}|m{2.9cm}|m{3.7cm}|m{1.8cm}|m{1.6cm}}
    \hline
     \multirow{2}{3.6cm}{material}    & ion diffusivity & electronic conductivity &   $\tau_D$   & $e_0$   \\ 
      &  [$\mathrm{m^2\ s^{-1}}$]  & $[\mathrm{S\ cm^{-1}}]$  & [s] & [$\mathrm{\mu J\ \mu m^{-2}}$] \\ \hline
      
     \multirow{2}{3.2cm}{Li$_{4+3x}$Ti$_5$O$_{12}$ (LTO) \\  $x:0$-1, i-m}  &  $10^{-16}$ \textsuperscript{[this work, $D_p^0$]} & {\scriptsize{$x=1$:}} 0.0035 \textsuperscript{[this work]}  & 1   & 0.055  \\ \cline{2-5}
     
     & $10^{-16}$-$10^{-15}$ \textsuperscript{\cite{Nitta2015Li-ionFuture, Wagemaker2009Li-ion4+xTi5O12}}  & {\scriptsize{$x=0$:}} $10^{-13}$-$10^{-6}$ \textsuperscript{\cite{Young2013ElectronicAnode}} \newline {\scriptsize{$x=1$:}} $10^{-2}$-1 \textsuperscript{\cite{ Zhao2015APerspectives, Young2013ElectronicAnode}}   & $10^{-1}$-1  & $10^{-2}$-$10^{1}$  \\ \hline

     Li$_{x}$CoO$_2$ (LCO) \newline $x:0.5$-1, m-i & $10^{-14}$-$10^{-12}$ \textsuperscript{\cite{Park2010ABatteries}} &  {\scriptsize{$x=0.5$:}} $10^2$ \textsuperscript{\cite{Milewska2014TheOxide}} \newline {\scriptsize{$x=1$:}} $10^{-4}$-$10^{-3}$ \textsuperscript{\cite{Milewska2014TheOxide}}  & $10^{-4}$-$10^{-2}$  & $10^{-1}$-$10^{1}$ \\ \hline
     
     Li$_x$MoS$_2$ (MoS$_2$) \newline $x:0$-$1$, i-m &
     $10^{-15}$-$10^{-14}$ \textsuperscript{\cite{Chen2018AtomicInterface}} & {\scriptsize{$x=0$:}} $10^{-7}$ \textsuperscript{\cite{Coleman2011Two-dimensionalMaterials, Wan2015InBatteries}} \newline {\scriptsize{$x=1$:}} $10^{-5}$-$10^{-4}$ \textsuperscript{\cite{ Wan2015InBatteries, Xiong2015LiProperties}}   & $10^{-2}$-$10^{-1}$  & $10^{-6}$-$10^{-4}$\\
    \hline
    
     Li$_x$FePO$_4$ (LFP) \newline $x:0$-1, i-i & 10$^{-15}$-$10^{-12}$ \textsuperscript{\cite{Malik2010ParticleDiffusivity, Park2010ABatteries}}  & {\scriptsize{$x=0.9$:}} $10^{-2}$ \textsuperscript{\cite{Kim2006EffectBatteries}}\newline {\scriptsize{$x=0,1$:}} $10^{-11}$-$10^{-10}$ \textsuperscript{\cite{Zhu2011ElectronicFePO4}}   & $10^{-4}$-$10^{-1}$  & $10^{-4}$-$10^{-1}$  \\ \hline
     
      Li$_x$C$_6$ (graphite) \newline $x:0$-1 (4 stages), m-m &  $10^{-15}$-$10^{-11}$ \textsuperscript{\cite{Nitta2015Li-ionFuture}} & {\scriptsize{$x=0$:}} $10^4$ \textsuperscript{\cite{Dresselhaus2002IntercalationGraphite}} \newline {\scriptsize{$x=0.25$:}} $10^5$ \textsuperscript{\cite{Dresselhaus2002IntercalationGraphite}}  & $10^{-5}$-$10^{-1}$  & 10$^1$-$10^5$  \\ \hline
     
    \end{tabular}
    \label{tab:materials}
\end{table*}

Next, we want to compare this work with our previous 1D model \cite{Tian2022InterfacialNanofilms} in terms of the analysis of LTO memristors. The old 1D model, with neutral wetting and no surface charge or surface energy relaxation,  has already made a reasonable explanation of the LTO memristor experiments \cite{Gonzalez-Rosillo2020Lithium-BatterySeparation}. It shows that the ion-poor system switches faster and requires smaller current, which is qualitatively consistent with experiments. However, it overestimates the switching current by 2 to 3 orders of magnitude if we take material parameters from literature. In this work, the 2D phase pictures further show that the ion-rich system can require even larger current to squeeze the less-conductive ion-poor phase to fully cover the electrodes, in order to make significant resistance change. In addition, we show that   surface charge, surface heterogeneity, and non-neutral wetting can reduce the switching current by over one order of magnitude (\autoref{fig:checkboard_c0.1}(b), \autoref{fig:checkboard_c0.9}(b)). 

Finally, there is still no direct experimental observation of MP. 
%In Ref.\cite{Gonzalez-Rosillo2020Lithium-BatterySeparation}, the Raman spectroscopy showing concentration drift at electric currents may be an indirect evidence for IPC. 
In the future, we will seek for colloborators to do in-situ TEM \cite{Liu2012InNanostructures, McDowell2013InNanospheres} to show MP directly.

\subsection{Improvements and extensions  of the model}

The current model can be improved in several aspects. First, we model the surface charge by a linear capacitor with only electronic charge considered. This greatly simplifies the problem. In the future, the model can be enriched by including  the nonlinear profiles of both ions and electrons in the space charge, and the dependence of capacitance on concentration and overpotential \cite{Bazant2009NonlinearVoltages, Bazant2009TowardsSolutions, Kilic2007StericCharging}. Second, we use an empirical Butler-Volmer equation with series resistance to describe the electron transfer kinetics at the electrode surface. This can also be improved by considering the physical mechanisms for electron transfer, such as tunneling and Schottky barrier \cite{Sze2007PhysicsDevices, Tung2014TheHeight}. Finally, the model can also be improved by including temperature change and mechanical effects.  

The current model can also be extended for other materials and applications. For example, the model can also be extended to multi-stage phase-separating ion-intercalation materials, like graphite \cite{Dresselhaus2002IntercalationGraphite} and WO$_3$ \cite{Zhong1992LithiumLixWO3}. This should lead to multiple states switchable by MP without including bulk or surface heterogeneity. Besides, we can also modify the boundary conditions in order to analyze multiphase coupled ion-electron transport in other ion-intercalation memories like RBT. We also hope this work will inspire material scientists to explore the potential role of MP in other applications such as battery. %In addition, the model may be used to explore the applicability of Li recovery from Li$_x$FePO$_4$ by multiphase polarization. 

\subsection{Conclusion}
To summarize, in this work, we derive a phase-field model for a mixed ion-electron conductor sandwiched by ion-blocking electrodes, for which model we have included a comprehensive boundary condition to consider the surface effects of electron transfer kinetics, non-neutral wetting, surface energy relaxation, and surface charge.  Then we apply the model to the ion-intercalation material, and study the phenomenon of multiphase polarization (MP) driven by high electric currents and the resulting resistive switching (RS). We show that the surface heterogeneity, surface charge, and non-neutral wetting can reduce the switching current significantly, and show that the manipulation of a small amount of the more-conductive phase in less-conductive phase can be better for LTO-like materials. we also compare the physics and performance of MP with other non-volatile RS mechanisms, and show that MP-based memories require multiphase ion-intercalation materials with high ionic diffusivity, low electronic conductivity, and significant metal-insulator transition with concentration. The phenomenon of MP is a great example of the coupling of ion and electron transport, and the model can be extended in the future for other problems with similar physics.

%\newpage

\medskip
\textbf{Supporting Information} \par %Please delete the Suppporting Information statement if it is not applicable. Please supply Supporting Information in another file. Supporting information should not be provided in .tex format
Supporting Information is available from the Wiley Online Library or from the author.

% Acknowledgements
\medskip
\textbf{Acknowledgements} \par %delete if not applicable))
This work was supported by a grant from Ericsson. 
The authors would like to thank Danniel Cogswell, Michael Li,  and Huada Lian for helpful discussions.

% References
\medskip

% Use the following code if you wish to generate your bibliography with BibTeX;
% replace the string "MSP-template" below with the name(s) of
% the BibTeX data base(s) you want to use.
% The resulting bibliography-output (the content of the .bbl file)
% must be pasted back into this file before submission.
% Please also include your BibTeX data base file(s) in your submission
% so that we can re-run BibTeX if necessary.
%
%\bibliographystyle{MSP}
%\bibliography{MSP-template}

%\textbf{References}\\
\bibliographystyle{MSP}
\bibliography{electrochem.bib, memories.bib, shockED.bib}

% Figures/tables and captions
% Permission statements are required for all figures reproduced or adapted from previously published articles/sources. Please also ensure that all necessary permissions to reproduce images have been received
% Please remove these statements for original figures

% Please provide Biographies and photos for Essays, Feature Articles, Progress Reports, Reviews, and Perspectives for those authors who should be highlighted  
% These should be at most 100 words long
% For other article types this section can be removed
% Photographs should be 40mm broad and 50 mm high

%\begin{figure}
%  \includegraphics{bio-placeholder.jpg}
%  \caption*{Biography}
%\end{figure}

% Table of contents entry should be 50 - 60 words long
% Image should be 55 mm broad and 50 mm high or 110 mm broad and 20 mm high

\clearpage 

\begin{figure}
\textbf{Table of Contents}\\
\medskip
  \includegraphics[width = 0.6 \textwidth]{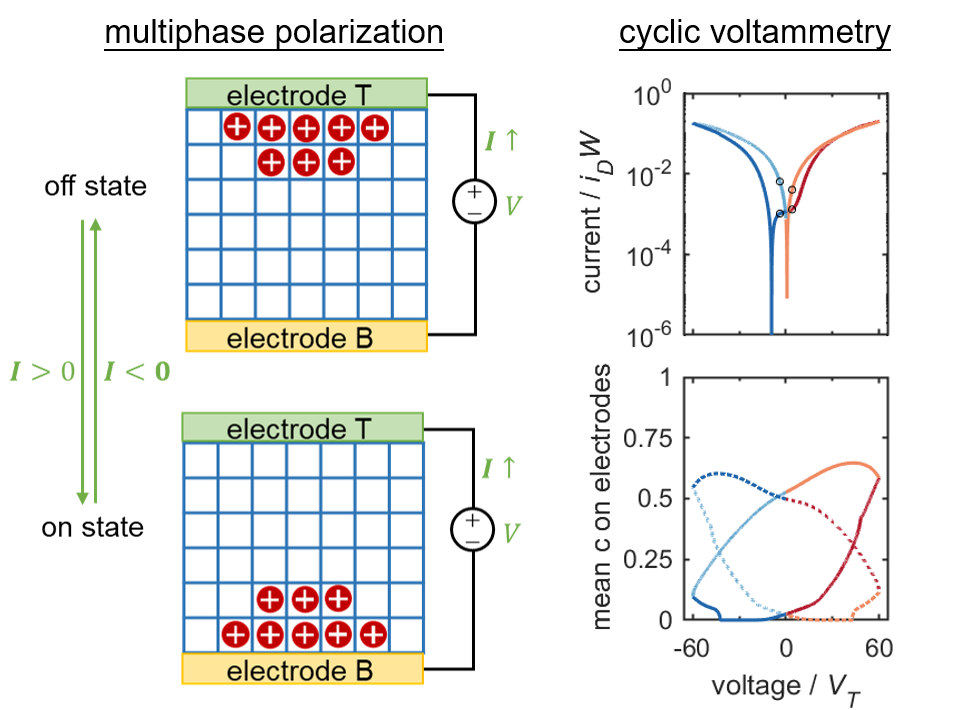}
  \medskip
  \caption*{ This work develops a comprehensive 2D phase-field model for coupled ion-electron transport in ion-intercalation materials, with surface effects including electron transfer kinetics, non-neutral wetting, energy relaxation, and surface charge. Then, the model is used to study the multiphase concentration polarization  in ion-intercalation materials between ion-blocking electrodes, and the corresponding resistive switching of the device.}
\end{figure}

\end{document}

\title{Supplemental Material: Multiphase  polarization in ion-intercalation nanofilms: general theory including various surface effects and memory applications}

\author{Huanhuan Tian}
\affiliation{Department of Chemical Engineering, Massachusetts Institute of Technology, Cambridge, MA, USA}

\author{Ju Li}
\affiliation{Department of Material Science and Engineering, Massachusetts Institute of Technology}
\affiliation{%
Department of Nuclear Engineering, Massachusetts Institude of Technology
}

\author{Martin Z. Bazant}
\email{bazant@mit.edu}
\affiliation{Department of Chemical Engineering, Massachusetts Institute of Technology, Cambridge, MA, USA}
\affiliation{Department of Mathematics, Massachusetts Institute of Technology, Cambridge, MA, USA}

\date{\today}

\renewcommand{\theequation}{S\arabic{equation}}
\renewcommand{\thefigure}{S\arabic{figure}}

\maketitle

\section{A. Derivation of Equation 10}{\label{appendix:app_1}}
In the following, we will use integration by parts:
\begin{equation}
    \int_{\mathbf{\Omega}} \nabla a \cdot \mathbf{b} dV =  \int_{\mathbf{\Omega}}  a \mathbf{n}\cdot \mathbf{b} dS - \int_{\mathbf{\Omega}} a \nabla \cdot \mathbf{b} dV
\end{equation}
where $a$ is any scalar and $\mathbf{b}$ is a vector. 

We first derive $\pd{\mathcal G_1}{t}$ :
\begin{equation}
\begin{aligned}
    \pd{\mathcal G_1}{t} =& \sum_{k=p,n}\int_{\mathbf{\Omega}} \left(\pd{g_k}{c_k} \pd{c_k}{t} + \pd{g_k}{\nabla c_k} \cdot \pd{\nabla c_k}{t} \right) dV \\
    &+\int_{\mathbf{\Omega}} \left( \sum_{k=p,n, d} \pd{g_k}{\phi} \right) \pd{\phi}{t} dV     \\
    &+ \int_{\partial \mathbf{\Omega}} \mu_n \mathbf{J}_n \cdot \mathbf{n} dS \\
    = & \sum_{k=p,n}\int_{\mathbf{\Omega}} \left(\pd{g_k}{c_k} - \nabla \cdot \pd{g_k}{\nabla c_k} \right) \pd{c_k}{t} dV \\
    & + \int_{\partial \mathbf{\Omega}} \pd{g_k}{\nabla c_k} \cdot \mathbf{n}\pd{c_k}{t} dS \\
    & +\int_{\mathbf{\Omega}} \left(  \sum_{k=p,n, d} z_k c_k F  \right)\pd{\phi}{t} dV \\
    & +  \int_{\partial \mathbf{\Omega}} \mu_n \mathbf{J}_n \cdot \mathbf{n} dS 
\end{aligned}
\label{eq:dG1dt_step1}
\end{equation}

Then we recognize that in the first integration, $\pd{g_k}{c_k} - \nabla \cdot \pd{g_k}{\nabla c_k} = \mu_k$, and $\pd{c_k}{t} = - \nabla \cdot \mathbf{J}_k$. Therefore,
\begin{equation}
\begin{aligned}
    & \int_{\mathbf{\Omega}} \left(\pd{g_k}{c_k} - \nabla \cdot \pd{g_k}{\nabla c_k} \right) \pd{c_k}{t} dV \\ 
    = & - \int_{\mathbf{\Omega}} \mu_k \nabla \cdot \mathbf{J}_k dV \\
    = & -  \int_{\mathbf{\partial \Omega}} \mu_k \mathbf{J}_k \cdot \mathbf{n} dS +  \int_{\mathbf{\Omega}} \nabla \mu_k \cdot \mathbf{J}_k dV
\end{aligned}
\end{equation}
Then we substitute this equation into \autoref{eq:dG1dt_step1}, and apply $ \partial g_k/\partial \nabla c_k = \kappa_k \nabla c_k$ and $\mathbf{J}_p \cdot \mathbf{n} = 0$ on $\partial \mathbf{\Omega}$, we can get Equation 10.

\section{B. Discussion on surface energy}

In this part, we first discuss on the signs of surface energy that we expect without mathematical model, based on physical insights and literature. And then we see if these expectations are consistent with our model.

To begin with, we compare the free energy of an electric double layer (EDL, two parallel layer of opposite charges) and a capacitor. The free energy of the natural formation of an  EDL, spontaneously formed at the interface between two materials without any externally applied voltage,  should be negative \cite{Verwey1948TheoryColloids}. This can be explained as below. The charge carriers in the two materials can have different chemical potential (Fermi energy at zero electric potential for electrons), so they tend to go from the high-energy side to the low-energy side, which will form the surface charges and develop an electric potential drop to balance the chemical energy drop. The above process occurs spontaneously, so the sign of the free energy should be negative. Although the electric double layer is often modeled as a capacitor, we should be careful about the differences and connections of those two. A capacitor can store energy so its own energy should be positive. However, the capacitor cannot be charged without external forces. A battery is needed to charge the capacitor. In another word, the free energy for an isolated circuit composed of a battery (an ideal voltage source in series with a resistor) in series with a capacitor should be negative. If we assume constant capacitance for the EDL, then the formula for the free energy of an isolated EDL is the same as that of an isolated battery-capacitor circuit.

Then let us see if our model is consistent with the above analysis.  Equation 16 (in combination with Equation 15) leads to
\begin{equation}
    \gamma_s = \gamma_s^* + \frac{1}{2} C\eta^2 - \frac{1}{2}C (\Delta \phi)_0^2.
\end{equation}
Therefore, $\eta = 0$ leads to a negative EDL energy, while $\Delta \phi_0 = 0$ leads to a positive capacitor energy. These are all consistent with our expectations. Then we can also check the second approach in Equation 18. We can change the limits of the integration, to get free energy for the formation of a natural EDL:
\begin{equation}
    \gamma_s^0 - \gamma_s^* = \int_0^{C (\Delta \phi)_0} \Delta \mu_n \frac{d \Gamma}{z_n F} = -\frac{1}{2} C(\Delta \phi)_0^2,
\end{equation}
which is as expected. This integral was actually used in Ref.\cite{Verwey1948TheoryColloids} to derive the free energy of an EDL. We can also see that Equation 18 is the energy stored in one capacitor due to external applied voltage. Finally, the energy into and out of the interface has been considered in Equation 4, and can change the effective surface energy, which has been shown in Equation 19.

\newpage
\section{C. supplementing figures}
\begin{figure}[hbt!]
    \centering
    \includegraphics[width = 0.8 \textwidth]{sub_figs/fig_sub_fun.png}
    \caption{The functions of $\tilde{\mu}_h(\tilde{c})$, $\tilde{\mathcal{I}}(\tilde{c})$, and $\tilde{i}_c^{ana}(\tilde{c})$ for the regular solution model with two sets of typical parameters. In the left column, the circles label the binodal points and the triangles label the spinodal points. } 
    \label{fig:my_label}
\end{figure}
\begin{figure}[hbt!]
    \centering
    \includegraphics[width = 0.8 \textwidth]{sub_figs/fig_step_c_0.1.png}
    \caption{Time evolution of the mean concentration on electrodes in the ion-poor systems, for all the cases shown in Figure 4.   Parts of the relaxation periods before and after the applied current  are also plotted here.}
    \label{fig:c_step_0.9}
\end{figure}

\begin{figure}[hbt!]
    \centering
    \includegraphics[width = 0.8 \textwidth]{sub_figs/fig_step_c_0.1_phi.png}
    \caption{Time evolution of the voltage applied to the ion-poor systems during the current step tests, for all the cases shown in Figure 4.   Parts of the relaxation periods before and after the applied current are also plotted here.}
    \label{fig:c_step_0.9}
\end{figure}

\begin{figure}[hbt!]
    \centering
    \includegraphics[width = 0.8 \textwidth]{sub_figs/fig_step_c_0.9.png}
    \caption{Time evolution of the mean concentration on electrodes in the ion-rich system, for all the cases shown in Figure 7.   Parts of the relaxation periods before and after the applied current are also plotted here.}
    \label{fig:c_step_0.9}
\end{figure}

\begin{figure}[hbt!]
    \centering
    \includegraphics[width = 0.8 \textwidth]{sub_figs/fig_step_c_0.9_phi.png}
    \caption{Time evolution of the voltage applied to the ion-rich systems during the current step tests, for all the cases shown in Figure 7.   Parts of the relaxation periods before and after the applied current  are also plotted here.}
    \label{fig:c_step_0.9}
\end{figure}

\bibliography{electrochem.bib, memories.bib}